\documentstyle[11pt,aaspp4,flushrt]{article}

\newcommand{\pone}{Paper~I}

\newcommand{\mum}{$\mu$m}
\newcommand{\cm}{cm$^{-3}$}

\newcommand{\hii}{H{\sc II}}
\newcommand{\uchii}{UC~H{\sc II}}

\newcommand{\nee}{$n_{\rm e}$}
\newcommand{\nc}{$N'_{\rm c}$}

\newcommand{\taucoo}{$\tau^{13}$}

\newcommand{\kms}{km~s$^{-1}$}
\newcommand{\vlsr}{$v_{\rm LSR}$}
\newcommand\hrrl{{H76$\alpha$}}
\newcommand\co{$^{12}$CO}

\newcommand\coo{$^{13}$CO}
\newcommand\cooj{$^{13}$CO J=1$-$0}
\newcommand\css{C$^{34}$S}
\newcommand\csj{CS J=2$-$1}

\newcommand\Ncoo{{$N(^{13}{\rm CO})$}}
\newcommand\Ncs{{$N({\rm CS})$}}

\newcommand{\Nhtwo}{$N({\rm H_2})$}
\newcommand{\nhtwo}{$n({\rm H_2})$}

\newcommand{\mlte}{$M_{\rm LTE}$}

\newcommand{\lir}{$L_{\rm IR}$}
\newcommand{\msol}{$M_\odot$}
\newcommand{\lsol}{$L_\odot$}
\newcommand{\ta}{$T_{\rm A}^*$}
\newcommand{\tr}{$T_{\rm R}^*$}
\newcommand{\tb}{$T_{\rm b}$}
\newcommand\tex{{$T_{\rm ex}$}}
\newcommand{\delv}{$\Delta v$}

\begin{document}


\title{
Molecular Counterparts of Ultracompact \hii\ Regions with Extended Envelopes
}

\author{Kee-Tae Kim\altaffilmark{1, 2} and Bon-Chul Koo\altaffilmark{2}}

\altaffiltext{1}{Department of Astronomy, University of Illinois,
1002 West Green Street, Urbana, IL 61801; ktkim@astro.uiuc.edu}
\altaffiltext{2}{Astronomy Program, SEES,
Seoul National University, Seoul 151-742, Korea; koo@astrohi.snu.ac.kr}


\begin{abstract}

We have carried out $^{13}$CO J=1$-$0, CS, and C$^{34}$S J=2$-$1
and J=3$-$2 line observations of molecular clouds
associated with 16 ultracompact (UC) \hii\ regions with extended envelopes.
The molecular clouds are the ones that 
give birth to rich stellar clusters and/or very massive (O7$-$O4) stars.
Our data show that the clouds are
very clumpy and of irregular morphology.
They usually have much larger masses, velocity dispersions,
and fractions of dense gas 
than molecular clouds that form early B or late O stars.
This is compatible with earlier findings 
that more massive stars form in more massive cores.
The IR luminosity-to-mass ratio has a mean value of 9~\lsol/\msol\
and is little correlated with the cloud mass.
Most molecular clouds have star formation efficiencies (SFE's) of 
1\%$-$2\%.
We find size-linewidth and size-density relations in the forms of
$\Delta v \propto D^{0.4}$ and $n({\rm H_2}) \propto D^{-1.2}$.
\coo\ cores are in general 
associated with compact \hii\ regions regardless of the presence of
\uchii\ regions therein. In contrast, CS cores are preferentially 
associated with compact \hii\ regions that contain \uchii\ regions. 
As with the fact that
the compact \hii\ regions containing \uchii\ regions are more compact 
than those not associated with \uchii\ regions, 
these indicate that
the former may be in an earlier evolutionary phase than the latter.
The diffuse extended envelopes of \hii\ regions often develop in 
the direction of decreasing molecular gas density.
Based on detailed comparison of molecular line data with radio continuum
and recombination line data,
the extended ionized envelopes are likely the results of champagne flows
in at least 10 sources in our sample.
Together these results appear to support a published suggestion
that the extended emission around \uchii\ regions can be naturally 
understood by combining the champagne flow model with 
the hierarchical structure of molecular clouds,
taking into account various inclinations and low resolutions of our data.
Additionally the blister model seems to be still applicable to 
most \hii\ regions,
even though massive stars usually form in the interiors rather than
on the surfaces of molecular clouds. 
It is possible because massive star-forming clouds have hierarchical
structure and irregular morphology.

\end{abstract}


\clearpage


\section{Introduction}

Ultracompact (UC) HII regions are very small (D$\lesssim$0.1~pc),
dense (\nee$>$10$^4$~cm$^{-3}$), and
bright (EM$>$10$^6$~pc~cm$^{-6}$) photoionized gas regions
(Wood $\&$ Churchwell 1989a; Kurtz, Churchwell, $\&$ Wood 1994).
They are thought to represent an early stage of massive-star evolution,
such that the newly formed stars are still embedded in their
natal molecular clouds.
Recently, however, it has been found that most \uchii\ regions are not
isolated but surrounded by diffuse extended emission
(Koo et al. 1996; Kim \& Koo 1996, 2001; Kurtz et al. 1999).
In Kim \& Koo (2001, hereafter \pone), we did a radio continuum and
radio recombination line (RRL) survey towards 16 \uchii\ regions 
with simple morphology and large ratios of single-dish to VLA fluxes,
and detected extended emission towards all the sources. 
The extended emission consists of 
one to several compact ($\sim$1$'$ or 0.5$-$5 pc) components and 
diffuse extended (2$'$$-$12$'$ or 4$-$19 pc) envelope.
All the \uchii\ regions but 2 spherical ones are located 
in the compact components,
where they always correspond to the peaks of their associated
compact components. 
(This positional coincidence we refer to as 
`the compact components with \uchii\ regions' or
`the compact \hii\ regions with \uchii\ regions' 
throughout this paper.)
Moreover,
the compact components with \uchii\ regions are smaller and denser than
those without \uchii\ regions.
Based on these morphological relations and RRL observations,
it seems certain that the ultracompact, compact, and extended
components in each source are physically associated,
and it is likely
that they are excited by the same ionizing source.
Thus almost all (14/16) the \uchii\ regions may not be 
bona-fide ionization-bounded `\uchii\ regions'
but rather ultracompact cores of more extended \hii\ regions.
A simple statistical analysis of single-dish to VLA flux ratios further 
suggested that most \uchii\ regions previously known 
are associated with extended emission like our sources.
Therefore, it is necessary to explain how the ultracompact, compact, and
extended components, which have very different dynamical time scales,
can coexist.

In \pone, we proposed a model in which the coexistence of 
the three components can be easily explained by combining the champagne flow 
model with the hierarchical structure of molecular clouds.
Recent high-resolution
molecular line observations of massive star-forming regions
revealed molecular clumps ($\sim$1~pc) and hot cores ($\lesssim$0.1~pc)
therein (see Kurtz et al. 2000 and references therein).   
The hot cores are likely to be the sites of massive star formation.
An interesting feature is that typical sizes of the hot cores and 
molecular clumps are, respectively, in rough agreement with
those of \uchii\ regions and their associated compact components.
Our model suggested, on the basis of these results, that 
when an O star formed off-center within a hot core embedded in a
molecular clump, 
the \hii\ region would continue to be ultracompact inside
the hot core for a long ($>$10$^5$~yr) time
while it grows to a few 1 pc outside the hot core.
In the light of the model, the coexistence of ultracompact, compact, and
extended components does not seem to be unusual for \hii\ regions that
expand in hierarchically clumpy molecular clouds, 
and the ultracompact and compact components would disappear one by one
when the host hot core and then molecular clump are destroyed
by the exciting star(s).
In the past decade a number of groups have undertaken studies
on hot cores and molecular clumps associated with \uchii\ regions 
(e.g., Cesaroni et al. 1994; Garay \& Rodr{\'\i}guez 1990).
However, there has been no holistic study of the relationship between 
the three components of \hii\ regions and their molecular gas
counterparts.

In this study, we made \coo, CS, and \css\ line observations
of molecular clouds associated with the 16 \uchii\ regions with extended
envelopes in \pone.
The main scientific goals of our observations are 
(1) to explore the physical characteristics of
the molecular clouds undergoing active massive star formation,
(2) to investigate how the HII region complexes interact with their parental
molecular clouds on diverse length scales, and 
(3) to understand the origin of the extended envelopes of \uchii\ regions.
The observational details are described in \S~2 
and the results are presented in \S~3. 
We discuss the physical properties and star formation of the molecular
clouds, the interaction between the ionized and molecular gas,
the origin of the extended halos of \uchii\ regions, and
the implication of our results for the blister model of \hii\ regions in \S~4.
Some individual sources are discussed in \S~5. 
We conclude with a summary of our main results in the last section.

\section{Observations}

\subsection{\cooj\ Line}

We have carried out \cooj\ (110.20135~GHz) line observations of 
molecular clouds associated with the 16 \uchii\ regions 
studied in \pone\
using the 12 m telescope of 
the National Radio Astronomy Observatory\footnote{
The National Radio Astronomy Observatory is 
operated by Associated Universities, Inc., under cooperative 
agreement with the National Science Foundation
} (NRAO) at Kitt Peak in 1997 June. 
The telescope has a full width at half-maximum (FWHM) of 60$''$
at 110~GHz.
The spectrometer used is a 256 channel filterbank with 64~MHz bandwidth
providing a velocity resolution of 0.68~\kms.
The system temperature was in the range 250$-$350~K.
All the molecular clouds were mapped with 20$''$ spacing
using the ``on-the-fly" observing technique.
We used positions with some emission as reference positions,
since we could not find any positions devoid of \coo\ emission near the sources.
We obtained a sensitive spectrum of each reference position and added
it to the position-switched spectra.
The observed temperature (\tr) was converted to main-beam brightness 
temperature (\tb) using the corrected main-beam efficiency (0.84)
provided by the NRAO. 
Table 1 shows the observational parameters.

\subsection{CS and \css\ Line}

We have conducted \csj\ (97.98095~GHz) line observations of 
the molecular clouds in three observational runs from 1996 November 
to 1997 April.
Ten sources were mapped with 60$''$ spacing.
In the remaining sources \csj\ emission was usually detected only
at the positions of the \uchii\ regions.
We have also observed CS J=3$-$2 (146.96904~GHz),
\css\ J=2$-$1 (96.41298~GHz), and \css\ J=3$-$2 (144.61711~GHz)
emission towards the \uchii\ regions in 2000 March.  
The observations were made with 
the 14 m telescope of the Taeduk Radio Astronomy Observatory (TRAO),
FWHMs of which are 60$''$ at 98 GHz and 46$''$ at 147 GHz.
The line intensity was obtained on the \ta\ scale
from chopper-wheel calibration.
The telescope has main beam efficiencies of 0.48 (98 GHz) and 0.39 (147 GHz).
All observations were made in the position switching mode
with reference positions checked to be free from 
appreciable CS emission (Table 3).
We used SIS mixer receivers with 256 channel filterbanks of 64~MHz 
bandwidth as the backend. The velocity resolutions
were 0.77~\kms\ and 0.51~\kms\ at 98 GHz and 147 GHz, respectively.
The system temperatures were typically 400K at 98~GHz and 700~K at 147~GHz.

\section{Results}

\subsection{\cooj\ Line Results}

Figures $1a-1p$ show our \coo\ spectra,
along with CS and C$^{34}$S spectra, at the positions of
all the 16 \uchii\ regions.  
There are in general two or more velocity components in the line of sight,
since the sources are located in the inner Galactic plane.
In each source, however, it is not difficult to find the velocity component
associated with \hrrl\ RRL emission, the center velocity of 
which is depicted by the vertical dotted line (\pone). 
The \hrrl\ line
is significantly ($\gtrsim$5~\kms) shifted from the associated \coo\ line
in 6 sources: G5.97$-$1.17, G10.15$-$0.34,
G10.30$-$0.15, G12.21$-$0.10, G23.71+0.17, and G27.28+0.15. 
The mean difference between the two velocities is 3.2$\pm$3.2~\kms\
for all the sources but G12.43$-$0.05 where \hrrl\ line emission was not
detected. 
A similar result was obtained by Forster et al. (1990),
who observed two hydrogen RRLs and seven molecular
lines towards 9 compact \hii\ regions 
and found velocity differences of typically 6~\kms. 
Thus such a velocity difference between the ionized and molecular gas
seems to be a common feature.
The velocity offset can be attributed to
bulk motion of the ionized gas as the \hii\ region expands asymmetrically 
in the direction of decreasing molecular density, 
as suggested by Forster et al. (1990).

Figures $2a - 2o$ exhibit integrated \coo\ line intensity maps of
molecular clouds.
The integrated velocity range is given at the top in each panel.
The \uchii\ regions and strong compact \hii\ regions are, respectively,
marked by large and small crosses in the individual panels
(see Table 3 of \pone).
The maps well reveal
clumpy structure and irregular morphology of molecular clouds,
especially for clouds at relatively small distances,
such as G5.89$-$0.39, G5.97$-$1.17, G8.14+0.23, G10.15$-$0.34,
and G10.30$-$0.15.
The typical size of clumps is $\sim$1$'$.
The UC and compact \hii\ regions are usually associated with dense
cores in the molecular clouds.
The individual sources will be discussed in some detail in \S~5.

If molecular gas is in local thermodynamical equilibrium (LTE),
the optical depth of \cooj\ line can be calculated from the observed 
intensity as compared with the excitation temperature which is assumed 
to be equal to the observed \co\ brightness temperature. 
In that case the \coo\ column density is given by the relation 

\begin{equation}
N(^{13}{\rm CO}) = 
2.42 \times 10^{14}~\frac{T_{\rm ex}~\int \tau^{13} dv}
{1~-~{\rm exp}(-5.29/T_{\rm ex})}~~~~({\rm cm^{-2}}),
\end{equation}

\noindent
where \tex\ is the excitation temperature in K, 
\taucoo\ is the optical depth, and $v$ is the velocity in \kms. 
We estimated the masses of molecular clouds, \mlte, using
the distances given by Wood \& Churchwell (1989) (Table 1)
and the \co\ brightness temperatures obtained from 
the Massachusetts-Stony Brook  Galactic Plane Survey
(Sanders et al. 1986).
It was assumed in this calculation that 
the mean molecular weight is 2.3~$m_{\rm H}$ and
the ratio of \Ncoo/\Nhtwo\ is 2$\times$10$^{-6}$ (Dickman 1978).

Table 2 lists the physical parameters of molecular clouds,
such as geometric mean radius $R$, \mlte,  H$_2$ molecule number density
\nhtwo, 
and FWHM of the average spectrum of cloud \delv.
The molecular clouds have sizes of 7$-$48~pc, 
masses of (1$-$102)$\times$10$^4$~\msol,
densities of (1.6$-$10.3)$\times$10$^2$ \cm, and 
line widths (FWHMs) of 4.1$-$10.2 \kms.

\subsection{CS and \css\ Line Results}

We detected \csj\ emission in all sources observed (Fig. 1). 
The center velocity of CS gas is in good agreement with that of \coo\ gas. 
Unlike \coo\ emission,
there is a single velocity component in CS line emission 
along the line of sight for all the sources with one exception (G23.46$-$0.20) 
and it is always associated with the \uchii\ region (see also Fig. 3). 
Towards G23.46$-$0.20, a velocity component with \vlsr$\simeq$100~\kms\ 
was observed over the entire area mapped (Figs. 1$i$ \& 3$g$) and 
an additional component 
was detected at about 80~\kms\ in the southern dense core. 
For the determination of the CS column density, \Ncs, we used 
the following formula

\begin{equation}
N({\rm CS}) = 1.88 \times 10^{11}~
T_{\rm r}~ {\rm exp}(7.05/{T_{\rm r}})~
\int{{T_{\rm b}} dv}~~~~({\rm cm^{-2}}),
\end{equation}

\noindent
which is valid for the optically thin case.
Here $T_{\rm r}$ is the rotation temperature in K, 
\tb\ is the brightness temperature in K,
and $v$ is the velocity in \kms.
We assumed $T_{\rm r}$ to be 30~K based on
studies of dense molecular gas in massive star-forming regions 
(see, e.g., Snell et al. 1984; Linke \& Goldsmith 1980; 
Churchwell et al. 1990). 
The \Nhtwo\ was derived from \Ncs\ using 
a fractional CS abundance to H$_2$ of 
1$\times$10$^{-9}$ (Linke \& Goldsmith 1980; Frerking et al. 1980). 
Table 3 presents \csj\ line parameters
and \Ncs\ at the positions of \uchii\ regions.

We also detected CS J=3$-$2 emission in all the \uchii\ regions
but one (G12.43$-$0.05), and \css\ J=2$-$1 and J=3$-$2 emission
in several sources (Tables 3 \& 4). 
The optical depth of \css\ emission, $\tau_{\rm p}$,
can be computed from the peak brightness temperature ratio
by assuming that the excitation temperature and the beam filling factor
are identical in the CS and \css\ lines in each transition, using the formula
 
\begin{equation}
T_{\rm b}({\rm C^{34}S}) / T_{\rm b}({\rm CS}) =
\bigl[ 1 - {\rm exp}(-\tau_{\rm p}) \bigr] /
\bigl[1 - {\rm exp}(-\tau_{\rm p}r) \bigr].
\end{equation}
 
\noindent
Here $r$ is the CS to \css\ abundance ratio, which we assumed to be equal to
the terrestrial value (22.5).
The estimated $\tau_{\rm p}$ and \css\ line parameters are summarized
in Table 4.

Figure $3a-3j$ show \csj\ integrated intensity maps of the 10 sources
mapped. CS-emitting regions are fairly clumpy and of irregular
morphology. The \uchii\ regions are nearly always associated with 
the strongest CS cores in the individual sources (see \S~4.2).
We determined the physical parameters of the CS-emitting regions, 
including $R$, \mlte, \nhtwo, and \delv\ (Table 5). 
The CS-emitting regions have sizes of 3$-$27~pc, 
masses of (3$-$470)$\times$10$^3$~\msol,
densities of (0.9$-$5.9)$\times$10$^3$~\cm,
and line widths (FWHMs) of 3.4$-$10.0 \kms.

\subsection{Correlation among Physical Parameters}

Similar power-law correlations were found
between $D$ and \delv\ and between $D$ and \nhtwo\
both for one line from cloud to cloud and for various lines of a cloud.
This suggests that a widespread and fundamental process operates
in molecular clouds.
Figure 4 is a plot of \delv\ against $D$
for the molecular clouds in our sample.
A fairly good correlation exists between the two parameters.
Here open circles are data points derived from \coo\ data, whereas
filled circles are the ones measured using CS data.
We performed a least-squares fit to the data points and obtained
log~\delv\ = (0.35$\pm$0.06) log~$D$ + 0.4$\pm$0.1
with a linear correlation coefficient of 0.80.
Figure 5 displays \nhtwo\ versus $D$. There is also a strong correlation
between the two. A least-squares fit yields
log~\nhtwo\ = ($-$1.24$\pm$0.15)~log~$D$ + 4.4$\pm$0.2
with a correlation coefficient of $-$0.86.
The estimated slopes of the two relations are in reasonable agreement 
with the values determined for large samples of molecular clouds
and dense cores in the Galaxy.
For example, Larson (1981) obtained 
\nhtwo~$\propto$~$D^{-1.1}$ and 
$\Delta v \propto$~$D^{0.38}$ relationships for about 50 
molecular complexes, clouds, and clumps.
Myers (1983) and Dame et al. (1986) also found similar power-law relations 
for smaller clouds and cores and for larger complexes, respectively. 
Several explanations were proposed to understand these relations but 
it is beyond the scope of this paper to mention them in detail 
(see, e.g., Myers \& Goodman 1988).
In \pone\ we found a relation between electron density and size
of the form $n_{\rm e} \propto$~$D^{-1.0}$ for \hii\ regions in our sample,
and proposed that the relation reflects variation in the ambient 
molecular gas density. 
The observed $D$-\nhtwo\ relationship with a similar power-law index 
for the host clouds and cores
might be direct evidence supporting the idea.

\section{Discussion}

\subsection{Physical Characteristics and Star Formation}
 
The molecular clouds in our sample are associated with extended
radio \hii\ regions,
each of which consists of an \uchii\ region, compact component(s),
and diffuse extended envelope.
As mentioned in \S~1,
the \uchii\ regions and their associated compact components
are likely to be ionized by the same sources.
In case where there are two or more compact components, 
e.g., G29.96$-$0.02 (Fig. 8$m$),
the compact components may be produced by 
separate ionizing sources (\pone).
Assuming that the compact components are excited by single
zero-age main-sequence stars,
their ionizing sources are O7 to O4 stars (see Table 4 of \pone).
If a stellar cluster is responsible for the ionization, alternatively,
the most massive star in the cluster would be about two subclasses later than
the equivalent single-star spectral type (cf. Kurtz et al. 1994).
In either case the observed molecular clouds seem to be the ones that 
give birth to very massive stars.

We compare the physical properties of our molecular clouds
with those of molecular clouds with some indicators of massive star
formation. Some groups made CO and \cooj\ line observations of
nearby molecular clouds associated with Sharpless \hii\ regions
(Carpenter, Snell, \& Schloerb 1995a; Heyer, Carpenter, \& Ladd 1996).
These molecular clouds appear to form lower-mass stars than our
molecular clouds, since the associated \hii\ regions are usually excited 
by early B  or late O stars.
They have sizes of 10$-$50~pc, masses of
(1$-$7)$\times$10$^4$~\msol, and line widths (FWHMs) of 2$-$6~\kms.
Despite similar sizes the molecular clouds in our sample
have considerably larger masses and velocity dispersions. 
Such molecular clouds were also studied in \csj\ line emission
(Zinchenko et al. 1994; Carpenter, Snell, \& Schloerb 1995b).
Their CS-emitting regions have sizes of 1$-$5~pc, masses of
(1$-$140)$\times$10$^2$~\msol, and line widths (FWHMs) of 2$-$3~\kms.
The CS-emitting regions of our molecular clouds 
are significantly larger, more massive, and more turbulent.
In Table 5 we give the estimated fraction of the CS-emitting region to 
the whole cloud in mass and area of our sample.
The CS-emitting regions on average contain 45$\pm$18\% of the total mass
and occupy 20$\pm$8\% of the total area.
The measured fractions are much higher than estimates for the nearby
molecular clouds with Sharpless \hii\ regions.
For example, the mass and area fractions are, respectively, 2\% and 1\%
for the Gemini OB1 cloud complex that contains several Sharpless 
\hii\ regions (Carpenter et al. 1995b).
This result is compatible with the finding of Carpenter et al. (1995b)
that more luminous $IRAS$ sources are in general associated with more massive 
CS cores in the Gemini complex, which strongly implies 
that more massive stars form in more massive cores.
Similar trends have been observed in the L1630 molecular cloud
(Lada 1992) and the Rosette molecular cloud (Phelps \& Lada 1997) as well.
In these molecular clouds the embedded clusters are associated with 
the most massive cores. Thus it was suggested  
that both high gas density and high gas mass
are required for the formation of stellar clusters.
In view of this, it is not surprising that
our molecular clouds, which form rich stellar clusters and/or
very massive single stars, have very massive CS cores.
 
The ratio of infrared (IR) luminosity to mass is a good measure
of the massive star formation activity of molecular clouds.
We derive the ratio for the molecular clouds in our sample using $IRAS$
HIgh RESolution processing (HIRES) images.
The total IR luminosity between 1~\mum\ and 500~\mum, \lir, can be 
estimated from the 60~\mum\ and 100~\mum\ flux densities in Jy, 
$F_{60}$ and $F_{100}$, by the relation 
(Lonsdale et al. 1985; Lee et al. 1996)
 
\begin{equation}
L_{\rm IR} ~= ~0.394 ~R(\overline{T_d}, \beta)
~\bigl[~ F_{100} + 2.58~ F_{60} ~\bigr]~ d^2 ~~~~~(L_\odot).
\end{equation}

\noindent
Here $T_d$ is the 60/100~\mum\ color temperature in K,
$\beta$ is the index in the emissivity law,
$Q_{\rm abs} \sim \lambda^{- \beta}$,
and $d$ is the source distance in kpc.
The $R(\overline{T_d}, \beta)$ is the color correction factor that accounts 
for the flux radiated outside the 60~\mum\ and 100~\mum\ $IRAS$ bands.
Table 6 lists \lir\ and \lir/\mlte\ for the individual sources. 
In this calculation $\beta$ was assumed to be 1 (Hildebrand 1983).
The ratio \lir/\mlte\ has a mean value of 9.1$\pm$6.0~\lsol/\msol,
which is much greater than the average ratio (2.8~\lsol/\msol) of all 
molecular clouds in the inner Galactic plane
but is comparable to the median value (7~\lsol/\msol) for 
molecular cloud complexes with \hii\ regions (Scoville \& Good 1989).
Figure 6 compares the ratio with \mlte. There does not seem to be 
any apparent correlation between the two parameters. 
This matches with the results of 
Mooney \& Solomon (1988), Scoville \& Good (1989), 
and Carpenter, Snell, \& Schloerb (1990), who found no correlation 
over the mass range $10^2 - 10^7$~\msol.

We examine the relationship between stellar mass, $M_\ast$, and cloud mass,
as shown in the lower panel of Figure 7.
In this comparison $M_\ast$'s were derived from 
the Lyman continuum photon fluxes, \nc's, measured by \pone\
using $<M_*>/<N_c '> = 5.7 \times 10^{-47}$
\msol~(photons~s$^{-1}$)$^{-1}$
(McKee \& Williams 1997; see also Mezger, Smith, \& Churchwell 1974). 
We performed a least-squares fit to all the data points
and obtained
log~$M_*$ = (0.62$\pm$0.12) log~M$_{\rm LTE}$ + 0.1$\pm$0.6
with a correlation coefficient of 0.83.
The slope agrees within errors with that (0.5$\pm$0.2) found by 
Myers et al. (1986) for a sample of 54 molecular cloud complexes 
in the inner Galaxy.
We determine star formation efficiencies (SFE's),
${M_\ast}/{(M_\ast + M_{\rm cloud})}$ (Table 6).
Most clouds in our sample have SFE's comparable to the Galactic median value 
of 1\%$-$2\% (Myers et al. 1986; Leisawitz, Bash, \& Thaddeus 1989),
while G5.97$-$1.17 and G37.55$-$0.11 have very high SFE's ($>$5\%).
They would have been classified as the most efficient star-forming clouds
by Myers et al. (1986), who found
only 4 molecular clouds with SFE's$\ge$4\% (M16, M17, W49, and W51)
in their sample (cf. Koo 1999).
Since they are the clouds with the least masses,
however, the high SFE's can be due to the disruption of molecular gas 
by massive stars.
This appears to be true for G5.97$-$1.17 (see \S~5.2).
If the two clouds are excluded, a least-squares fit yields
log~$M_*$ = (0.86$\pm$0.12) log~M$_{\rm LTE}$ $-1.3\pm$0.7
with a higher (0.90) correlation coefficient,
so that the slope appears to be fairly steeper than the former.
The value of SFE has little correlation with \mlte,
as might be expected from the independence of \lir/\mlte\ on \mlte.

\subsection{Interaction between Ionized and Molecular Gas}

Figures 8$a$-8$o$ compare
the distribution of ionized gas with that of molecular gas in our sources.
The extended envelopes of \hii\ regions seem to be largely determined by 
the ambient molecular gas distribution, since they often develop 
in the direction of decreasing molecular gas density.
G8.14+0.23 and G10.30$-$0.15 provide good examples.
In these sources, the central compact components are associated with
\coo\ cores, while the diffuse envelopes stretch out oppositely along 
the axes approximately orthogonal to the elongated natal molecular clouds.
Such a bipolar morphology is very reminiscent of champagne flows
that would arise from a flat molecular cloud
(Bodenheimer, Tenorio-Tagle, \& Yorke 1979;
Tenorio-Tagle, Yorke, \& Bodenheimer 1979).
The large velocity gradients expected in the champagne flows were not 
observed in these objects, however, although they could be present
but masked by large inclination angles.
We have found using the \hrrl\ line data in 7 sources
(G5.89$-$0.39, G5.97$-$1.17, G10.15$-$0.34,
G10.30$-$0.15, G12.21$-$0.10, G23.46$-$0.20, and G29.96$-$0.02)
that the extended envelopes are blue- or red-shifted 
by 3$-$10~\kms\ from the \uchii\ regions (\pone).
The \hrrl\ line is considerably shifted with respect to molecular lines
in several sources, as noted in \S~3.1.
Based on the morphological comparison between \hii\ regions and molecular
clouds, the velocity gradient in \hrrl\ line emission, 
and the velocity difference between \hrrl\ and molecular lines,
the extended envelopes can be interpreted as consequences 
of champagne flows in at least 10 sources in our sample:
G5.97$-$1.17, G8.14+0.23, G10.15$-$0.34, G10.30$-$0.15, G12.21$-$0.10,
G12.43$-$0.05, G23.46$-$0.23, G23.71+0.17, G23.96+0.15, and G27.28+0.15
(Israel 1978; Tenorio-Tagle 1979; Franco, Tenorio-Tagle, \& Bodenheimer 1990).
It should be noted that 
an obvious anti-correlation between the distributions of the ionized and 
molecular gas cannot be seen unless the inclination of the champagne flow is 
large, and that
low resolution and coarse sampling of our RRL observations
prevent us from examining the internal velocity gradient of the ionized gas 
in the majority of our sources.
The results of this inspection on the individual sources will be discussed
in \S~5.

We investigate the association between compact \hii\ regions 
and \coo\ cores from Figure 8.
The vast majority of compact \hii\ regions are associated with \coo\ cores, 
which are in general the densest ones in the individual sources.
However, the radio continuum peaks and their associated \coo\ peaks 
usually do not overlap. This is more clearly shown in Figure 9, 
which is a histogram demonstrating the number of compact \hii\ regions at each
separation interval between the peaks of radio continuum and \coo\ intensities.
Figure 9 also displays that there may be no significant difference in 
statistics between the compact \hii\ regions with and without \uchii\ regions.
We performed a similar analysis for CS cores as well. 
All the compact HII regions with \uchii\ regions
are intimately related to CS peaks, 
whereas those without \uchii\ regions are mainly shifted from the peaks
(see, e.g., Figs. 3$d$ \& 3$f$).
CS gas was detected only at the positions of \uchii\ regions in the sources
not mapped, as said earlier.
We did not analyze statistically the separation of the radio continuum
peaks and the associated CS peaks,
because the sampling interval (60$''$) of our CS line maps is 
comparable to or larger than the typical angular separation between the
two peaks.

\subsection{Origin of Extended Emission around \uchii\ regions}

In \pone\ we suggested a simple model to explain the origin of the extended 
envelopes of \uchii\ regions (see also \S~1). In the model the extended envelopes
are produced by champagne flows originating from hot cores of molecular clouds. 
However, we could not completely exclude the possibility that
they are formed during the initial ionization of \hii\ regions. 
This is possible if hot cores and molecular clumps are so clumpy that 
ultraviolet (UV) photons can propagate a few 1~pc from the central stars. 
How can we distinguish between the two possibilities?
For doing this it seems to be important to compare in detail the ionized gas 
distribution in the envelopes with the ambient molecular gas distribution. 
The extended envelopes are expected to stretch out in the direction 
of decreasing molecular gas density in both cases.
If they were formed very soon after the central stars turned on,
the density distribution of ionized gas would be very similar to that of 
molecular gas because it is not yet significantly affected by 
dynamical evolution.
On the contrary, 
the two density distributions could be substantially different each other
if the extended envelopes were formed by champagne flows.
In this context, our data appear to support the champagne flow model.
In G8.14+0.23 at a distance of 4.2~kpc, for example, 
the ionized gas protuberances extend into molecular gas hollows 
that lie north and south of the central dense core 
(Fig. 8$d$).
We can also see similar features in other sources
at relatively small ($d$$\le$6~kpc) distances, 
such as the eastern part of G10.15$-$0.34 and 
the northern part of G10.30$-$0.15 (Fig. 8$e$).
In particular, a velocity variation of $\sim$10~\kms\ was actually
observed along $\delta (1950) \simeq -20^\circ 20' 05''$
in the extended envelope of G10.15$-$0.34 (Kim \& Koo 2002).
There is sometimes a significant velocity difference
between the RRL and molecular line emission observed towards each 
\uchii\ region.
The RRL width is generally larger in the compact \hii\ regions,
especially the ones with \uchii\ regions, than in the extended envelopes
(\pone; Garay \& Rodr{\'\i}guez 1983).
These might be also strong arguments in favor of the champagne flow model,
even though detailed comparison of radio continuum and RRL data
with molecular line data at a higher resolution 
is needed to clarify this issue.

We found in \pone\ that
the compact \hii\ regions with \uchii\ regions are smaller and denser than
those without \uchii\ regions,
and proposed that the former
might be in an earlier evolutionary phase than the latter.
Since massive stars emit strong UV radiation, stellar winds, and energetic 
bipolar outflows even at the earliest evolutionary stage,
the host dense cores should be disrupted with time.
Thus our suggestion matches with 
that the compact \hii\ regions with \uchii\ regions
are more closely associated with CS peaks than those without \uchii\ regions.
These observations are consistent with our model.
On the other hand,
two spherical \uchii\ regions in our sample, G23.46$-$0.23 and G25.72+0.05,
have no associated compact \hii\ regions (Figs. 8$h$ \& 8$k$).
G25.72+0.02 is coincident with a \coo\ core
and has detectable CS emission only at the \uchii\ region position,
but
G23.46$-$0.20 is significantly offset from the nearby \coo\ and CS peaks
(Figs. 2$h$ \& 3$g$). 
The absence of associated dense core and compact \hii\ region
indicates
that G23.46$-$0.23 may be an externally ionized dense clump.
In contrast, G25.72+0.05 is likely to be the only bina-fide 
ionization-bounded \uchii\ region
out of the 16 \uchii\ regions in our sample.\\

\subsection{Revisit to Blister Model of \hii\ regions}

Zuckerman (1973) for the first time suggested 
that the Orion Nebula is a blister-like \hii\ region located 
on the near edge of an associated molecular cloud.
After that
Israel (1978) found from a comparison of optical \hii\ regions and molecular
clouds that the majority of \hii\ regions in his sample have similar structures
to the Orion nebula and further proposed that the formation of massive stars
occur near the surfaces of molecular clouds (see also Israel 1976, 1977). 
Subsequent numerical simulations successfully demonstrated that
such a blister-like morphology of \hii\ regions can be produced by
champagne flows derived by density contrast between molecular clouds and
intercloud medium and/or strong negative density gradient in molecular clouds
(e.g., Tenorio-Tagle 1979; Franco et al. 1990).
However, Waller et al. (1987) showed
by comparing radio \hii\ regions and molecular clouds 
that radio \hii\ regions are preferentially concentrated toward 
the centers of molecular clouds,
and argued that the result conflicts with the blister picture.

On the other hand,
our data show that the compact (and ultracompact) components of 
\hii\ regions are closely associated with dense cores of molecular clouds. 
Since most of the dense cores are deeply embedded in molecular clouds,
it is likely that massive stars in general form 
in the interiors rather than on the surfaces of molecular clouds.
This is consistent with what Waller et al. (1987) found.
However, our data strongly suggest that the diffuse extended 
envelopes of \hii\ regions are produced by champagne flows.
Thus the blister picture appears to be still applicable to most HII regions.
It seems possible because massive star-forming molecular clouds
have hierarchical structure and are of irregular morphology.
In view of our model,
the peaks of optical \hii\ regions are likely to be located around
the edges of associated  molecular clouds 
not because the exciting star(s) formed there,
but because the side of the molecular cloud had already been disrupted 
by champagne flows.

\section{Comments on Some Individual Sources}

\subsection{G5.89$-$0.39}

This molecular cloud provides a good example showing that the molecular 
clouds in our sample are very clumpy and of irregular morphology (Fig.~2$a$).
The molecular cloud is moderately elongated in the northwest-southeast 
direction on the whole and is largely divided into three components: 
central, northwestern, and southeastern ones. 
Apparently massive star formation is ongoing only in the central component.
The central component has two dense cores, which are associated 
with the strong compact components in the radio continuum (Fig.~8$a$).
The \uchii\ region is located in the western denser core,
which is consistent with the idea that
the compact components with \uchii\ regions are in an earlier
evolutionary stage than those without \uchii\ regions.

\subsection{G5.97$-$1.17}

Among our sources this is the nearest (1.9~kpc) one and so 
the molecular cloud reveals very well the clumpy structure (Fig.~2$b$).
The cloud is composed of an extended ($\sim$6$'$$\times$10$'$) component 
and an arclike structure, which emanates from the former to the southeast.
The arclike structure has radio continuum counterpart (Fig. 8$b$).
The \uchii\ region is associated with a weak dense core, which
resides at the edge of the central component.
The strongest \coo\ core is located northwest of the cloud
but shows no indication of ongoing massive star formation
in radio and mid-infrared continuum emission (Crowther \& Conti 2003).
As pointed out in \pone, 
our extended radio \hii\ region is
the dense part of the well-known Lagoon nebula (M8).
It seems that
the dense part is ionized mainly by Herschel 36 (O7), nearly corresponding
to the \uchii\ region in position, while more extended optical nebulosity is 
excited by other O stars, including 9~Sgr (O4(f)) (Elliot et al. 1984).
The ionized gas is significantly ($>$5~\kms) blue-shifted from the
molecular gas (Fig. 1$b$ and Table 5 of \pone),
which indicates that the \hii\ region is on the front edge of the
host molecular cloud and the ionized gas is approaching the Earth 
with respect to the molecular cloud, as proposed by Lada et al. (1976).
The star 9 Sgr is located about 3$'$ east of \uchii\ region, i.e., 
$(\alpha, \delta)_{1950}$=
($18^{\rm h} 00^{\rm m} 49.^{\!\!\rm s}8$, $-24^\circ 22' 00''$).
Thus the eastern hollow may have been produced by strong UV radiation and
stellar winds from the star 9 Sgr. 
Consequently, the molecular cloud seems to have been considerably 
disrupted by O stars therein.  
This is compatible with the fact that
the molecular cloud has a much higher SFE, 7.0\%,
in comparison with other clouds in our sample (Table 6).

\subsection{G8.14+0.23}

The molecular cloud is largely extended in the northeast-southwest 
direction, but the central dense part is elongated 
in the east-west direction (Figs 2$d$ \& 3$c$).
The \uchii\ region is placed in the flattened dense core, 
a little west of the peak, which strongly supports 
that the bipolar morphology of \hii\ region is the result of 
champagne flows originating from a thin molecular cloud (see \S~4.2). 
In the view of the model presented in \pone, this object is
a good example of bipolar blister-type \hii\ regions observed 
at a viewing angle close to 90$^\circ$.  

\subsection{G10.15$-$0.34 and G10.30$-$0.15}

This region is known as the W31 \hii\ region/molecular cloud complex.
The molecular cloud seems to be an incomplete shell, being open to the
northwest, on the whole (Fig. 2$e$).
The two \uchii\ regions are located in the southern spherical dense 
region and in the northern flat one, respectively. 
We found convincing observational evidence that their extended ionized
envelopes have been produced by champagne flows (e.g., see Fig. 8$e$).  
A detailed radio and infrared study on this region was presented
in a separate paper (Kim \&\ Koo 2002).

\subsection{G12.43$-$0.05}

The \uchii\ region is situated outside the \coo\ core, although it has an 
associated compact component in the radio continuum (Fig. 8$g$).
We have detected weak (\ta$\simeq$0.3~K) \csj\ line emission only at 
the position of the \uchii\ region (Fig. 1$h$). 
The associated compact component is the largest and most diffuse among 
the 14 compact components with \uchii\ regions in our sample 
(see Tables 3 \& 4 of \pone). 
In the picture of our model, 
it may suggest that the natal hot core has been almost destroyed 
by the ionizing star(s).
Thus the compact component is likely 
on the verge of being one without ultracompact core.

\subsection{G23.46$-$0.20}

G23.46$-$0.20 is one of the two spherical \uchii\ regions that have no 
associated compact components in the radio continuum images. 
It is located near the boundary of extended envelope (Fig. 8$h$).
There are no associated \coo\ or CS cores as well (Figs. 2$h$ \& 3$g$).
Therefore, G23.46$-$0.20 might be a dense clump externally ionized 
by nearby luminous stars. 
The molecular cloud is roughly extended in the northwest-southeast direction,
while the diffuse envelope of \hii\ region is elongated 
in the east-west direction.
According to RRL data obtained in \pone,
the northern and western parts are blue-shifted by $\sim$4~\kms\
from the central region, which has approximately the same 
($\sim$104~\kms) center velocity as the molecular gas (Fig. 1$i$).
These observations may indicate that the extended envelope
is the result of champagne flows. 

\subsection{G23.71+0.17}

The \uchii\ region corresponds to the peak of an extended \hii\ region
whose compact component is associated with a \coo\ core (Fig 8$i$). 
CS J=2$-$1 and J=3$-$2 emission were detected only towards the \uchii\ 
region (Table 3). 
The \hrrl\ line (103.0~\kms) is blueshifted from the \coo\ and CS lines by about 
12~\kms\ (Fig 1$j$), suggesting that the \hii\ region may be located on 
the near edge of the molecular cloud.
The situation is very similar to G5.17$-$1.17 (\S~5.2) or the Orion nebula
(Zuckerman 1973),
even if this object is at a much larger (9~kpc) distance. 
The viewing angle does not seem to be very small because the 
diffuse envelope of the \hii\ region is more extended to the northeast.

\subsection{G23.96+0.15 and G27.28+0.15}

These two \uchii\ regions are located in \coo\ and CS cores.
The molecular clouds are elongated in the north-south direction,
while the diffuse envelopes of \hii\ regions fairly extended
in the perpendicular direction (Figs 8$j$ \& 8$l$).
Thus the extended envelopes can be interpreted as champagne flows. 
However, we could not observe any internal velocity gradient 
due to low (FWHM$\simeq$2$'$) angular resolution of RRL observations.

\subsection{G29.96$-$0.02}

The \uchii\ region is associated with a local \coo\ peak, whereas 
it is located in the strongest CS peak (Figs 2$m$ \& 3$j$). 
The dense region of the molecular cloud 
is somewhat elongated in the northeast-southwest direction. In the channel maps
of both \coo\ and CS line data, a significant ($\sim$8~\kms)
velocity gradient is present. Strong \coo- and CS-emitting regions 
move from northeast to southwest as the velocity increases
from \vlsr$\simeq$92~\kms\ to 100~\kms.  
We can find a similar velocity gradient in the RRLs 
taken in approximately full-beam spacing
along $\alpha$(1950) $\approx$ $18^{\rm h}43^{\rm m}28.\!\!^{\rm s}0$
(see Table 5 of \pone).

\section{Conclusions}

We carried out $^{13}$CO J=1$-$0, CS, and C$^{34}$S J=2$-$1
and J=3$-$2 line observations of the parental molecular clouds of
16 \uchii\ regions with extended envelopes, which give rise to 
rich stellar clusters and/or very massive (O7$-$O4) stars.
The molecular clouds in our sample are very clumpy and of irregular 
morphology. Our molecular clouds and dense cores usually have much 
larger masses and velocity dispersions than those 
associated with Sharpless \hii\ regions
that are ionized by early B or late O stars.
The CS-emitting regions on average contain $\sim$45\%
of the total cloud mass and occupy $\sim$20\% of the total area
for the 10 sources mapped in the CS J=2$-$1 line.
The estimated fractions are one order of magnitude higher than
the values for the molecular clouds that form early B or late O stars.
This result is consistent with the earlier findings for smaller dense cores
that more massive stars form in more massive cores.

\coo\ cores are in general associated with compact \hii\ regions
regardless of the presence of \uchii\ regions therein,
while CS cores are preferentially associated with those with
\uchii\ regions. 
Along with the fact that the compact \hii\ regions with \uchii\ regions
are more compact than those without \uchii\ regions,
this strongly suggests that
the former are in an earlier evolutionary stage than the latter.
The extended envelopes of \hii\ regions tend to develop 
towards molecular regions of low density.
It was found in some sources at relatively small ($d$$\le$6~kpc) distances
that the density distribution of ionized gas in the envelopes
is substantially different from that of the ambient molecular gas,
indicating that the ionized gas distribution have been significantly
affected by dynamical evolution.
Thus it seems likely that the extended envelopes have been formed 
by champagne flows rather than the initial ionization,
although higher-resolution radio continuum, molecular line, and especially 
RRL observations are required to clarify this issue.
Based on the morphological comparison of \hii\ regions and molecular clouds,
the velocity gradient in \hrrl\ line emission,
and the velocity difference between \hrrl\ and molecular lines,
the extended ionized envelopes are likely to be the consequences
of champagne flows in at least 10 sources in our sample.
Together, these results appear to support our model that explains 
the presence of low-density halos of \uchii\ regions 
by combining the champagne flow model with the hierarchical structure 
of molecular clouds,
taking into account various inclinations and low resolutions of our data.

Our results do not imply, however, that the formation of massive stars 
takes place near the surfaces of molecular clouds
as proposed by the original blister model of \hii\ regions (Israel 1978).
Instead, since the UC and compact components of \hii\ regions
mainly
correspond to the dense cores embedded in molecular clouds,
massive stars seem to form in the interiors of molecular clouds.
Nevertheless, 
the diffuse extended envelopes of \hii\ regions can be produced by
champagne flows
because massive star-forming molecular clouds have hierarchical structure
and irregular morphology.

\acknowledgements
We are very grateful to Ed Churchwell, Stan Kurtz, 
and Guillermo Tenorio-Tagle for thoughtful comments
and discussion. We also thank the anonymous referee 
for helpful suggestions and comments.
This work has been supported by
BK21 Program, Ministry of Education, Korea through SEES,
and also the Laboratory for Astronomical Imaging 
at the University of Illinois and NSF grant AST 99-81363.

\clearpage

\clearpage

\begin{figure}
\vskip 0cm
\figurenum{1}
\epsscale{1.0}
\plottwo{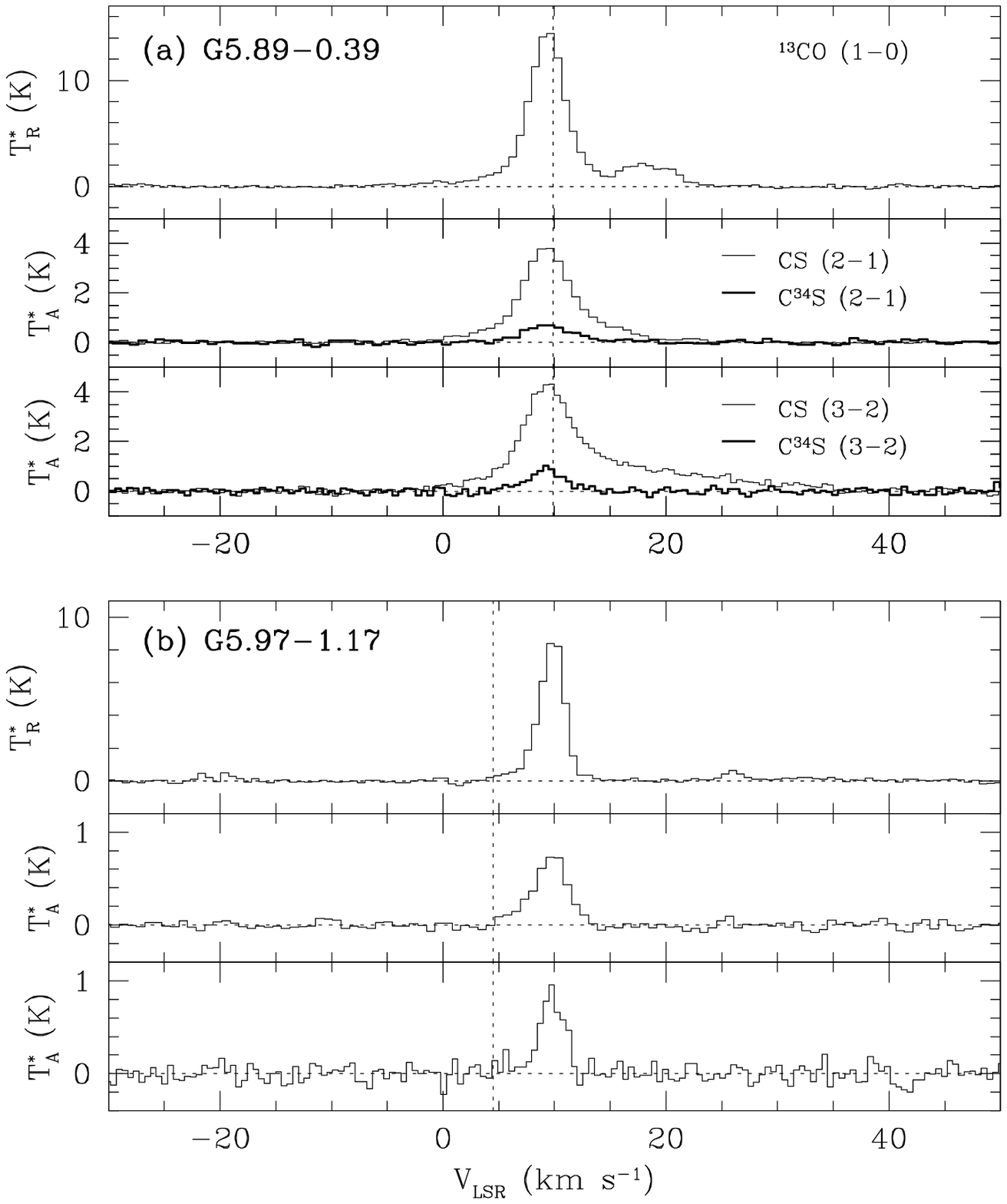}{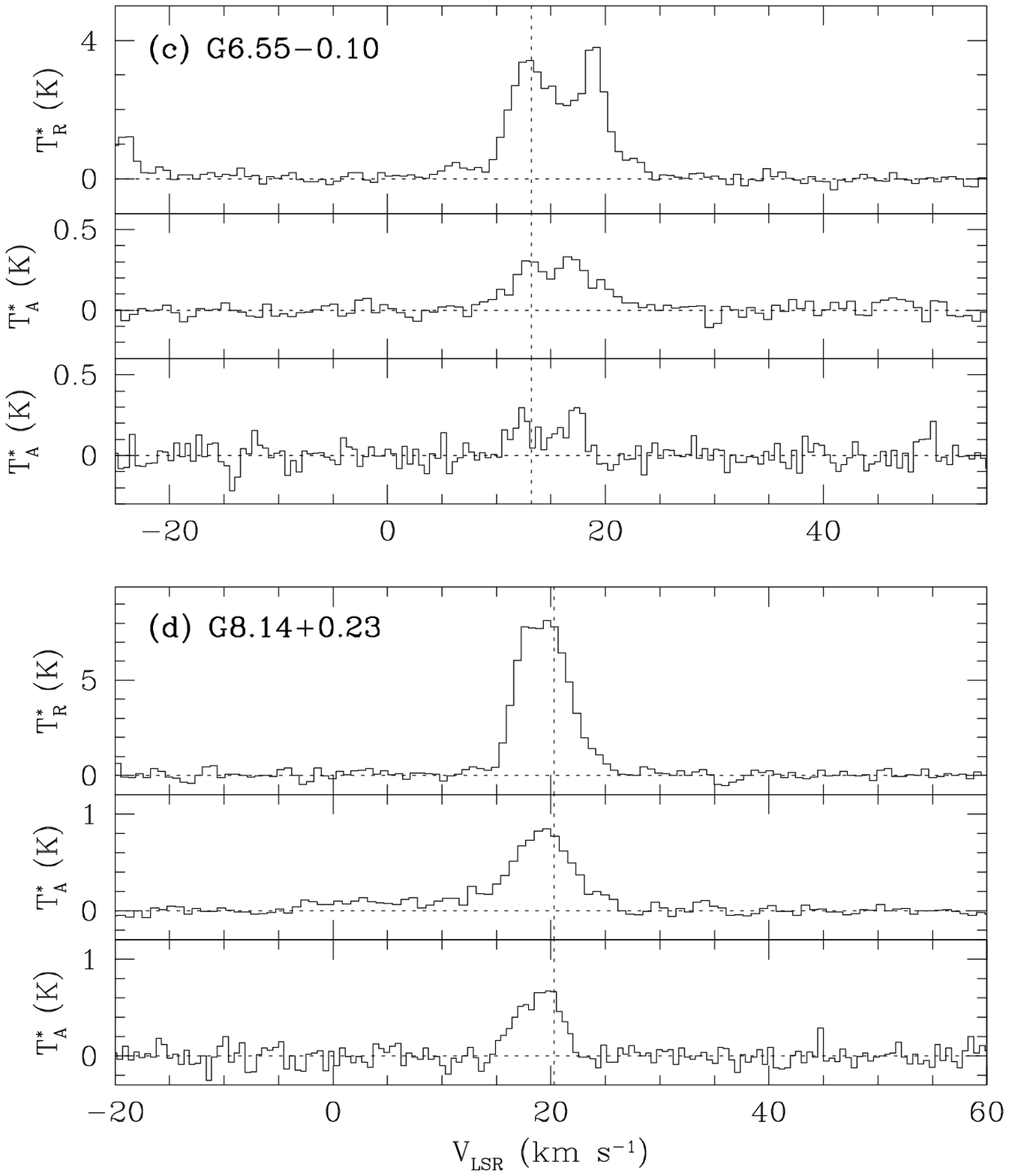}
\end{figure}

\begin{figure}
\vskip 0cm
\figurenum{1}
\epsscale{1.0}
\plottwo{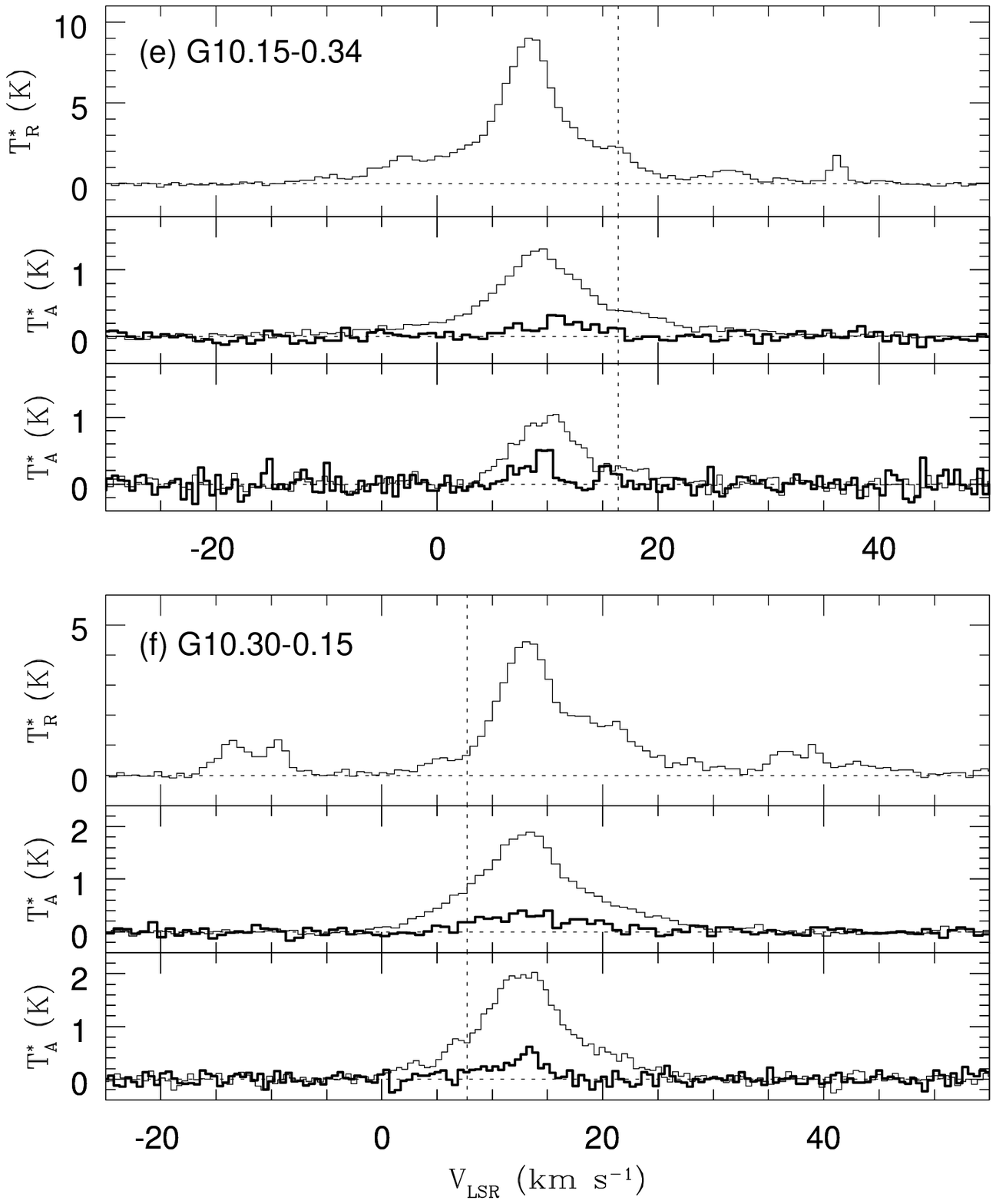}{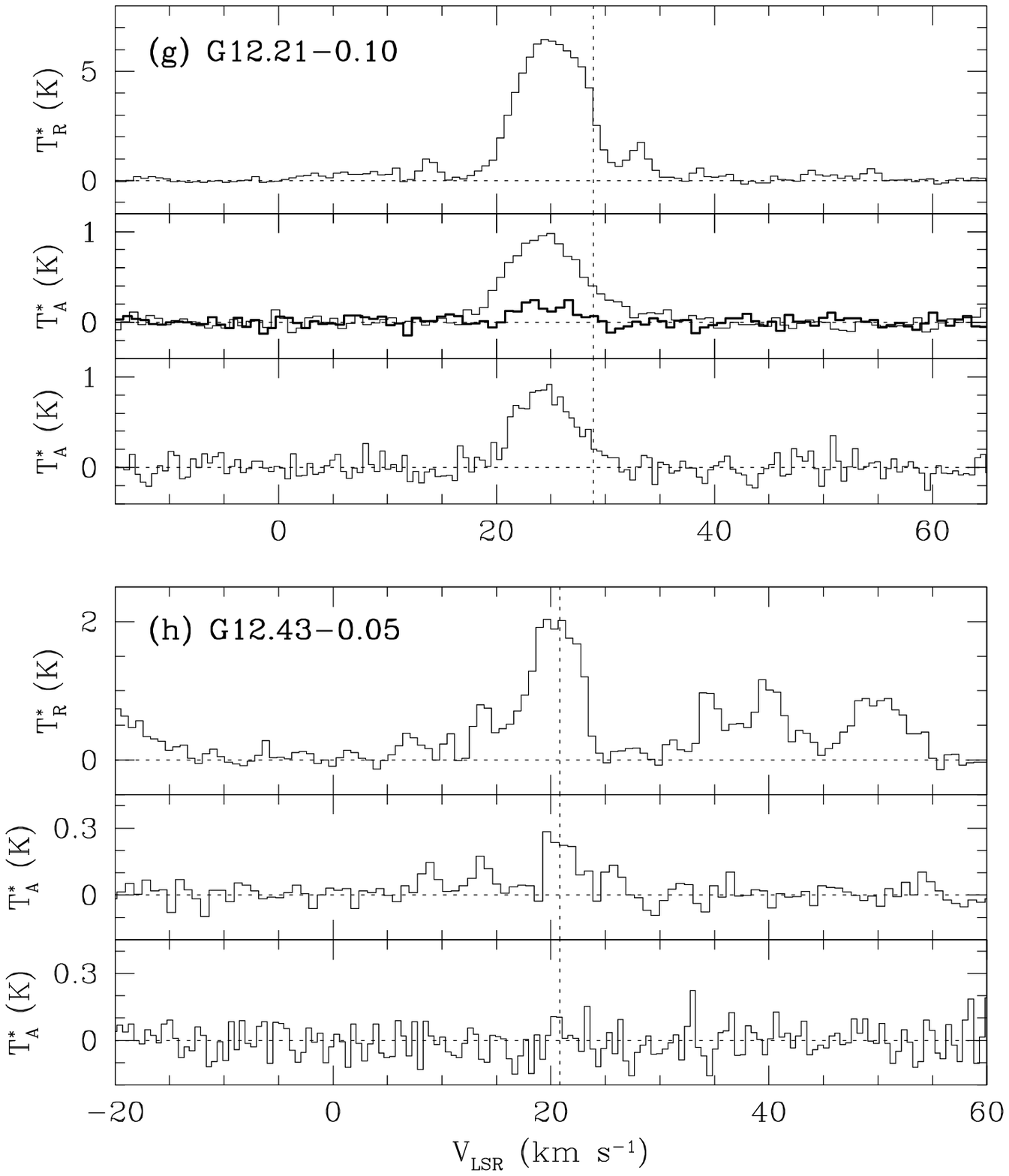}
\vskip 0cm
\figcaption[f1.eps]{
\coo, CS, and \css\ line profiles at the positions of \uchii\ regions.
In each panel, the vertical dotted line represents the center velocity
of \hrrl\ line emission obtained in \pone\ except for G12.43$-$0.05.
It indicates the center velocity of NH$_3$ (2,2) line in G12.43$-$0.05 (1$h$).
}
\end{figure}

\begin{figure}
\vskip 0cm
\figurenum{1}
\epsscale{1.0}
\plottwo{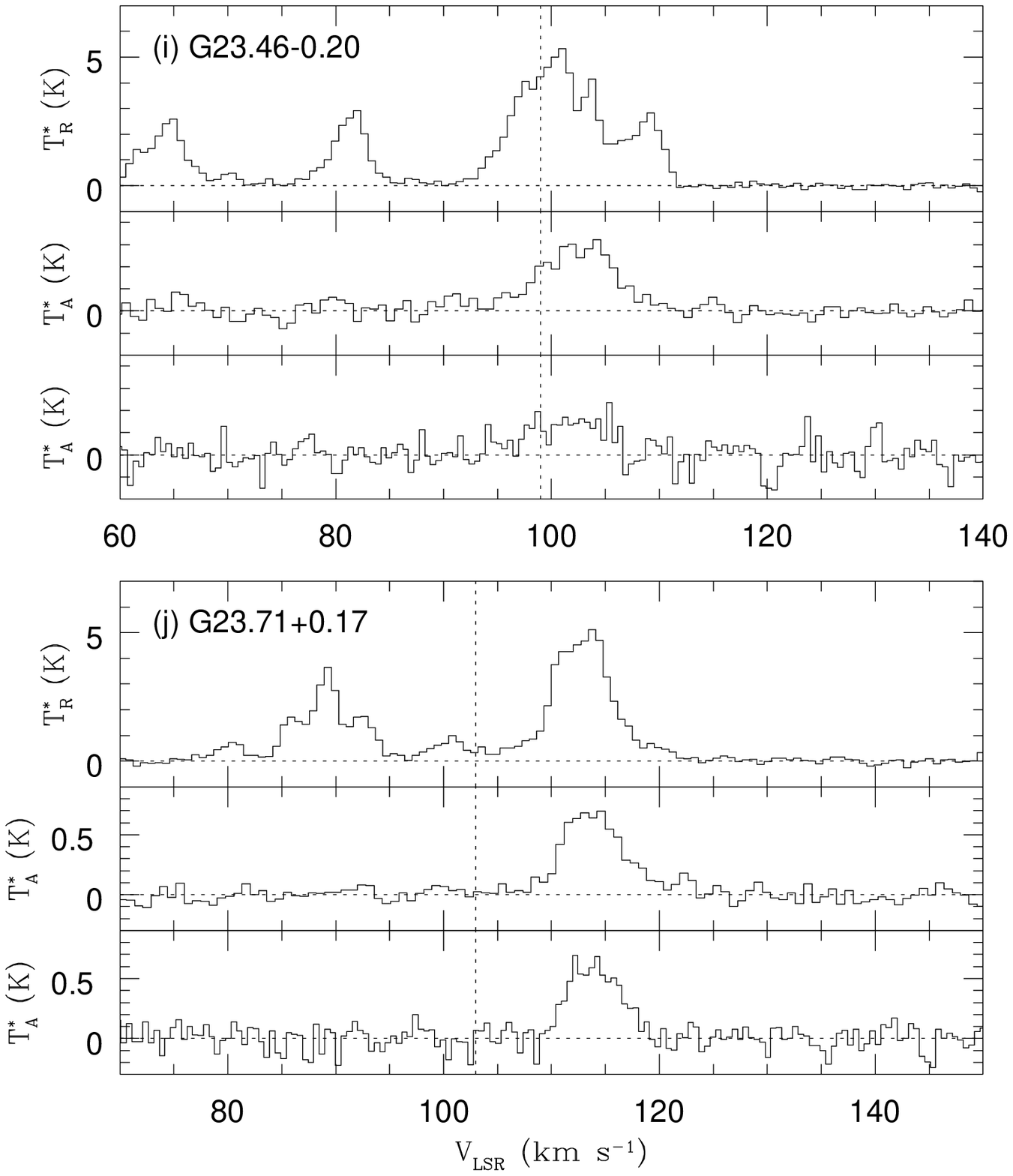}{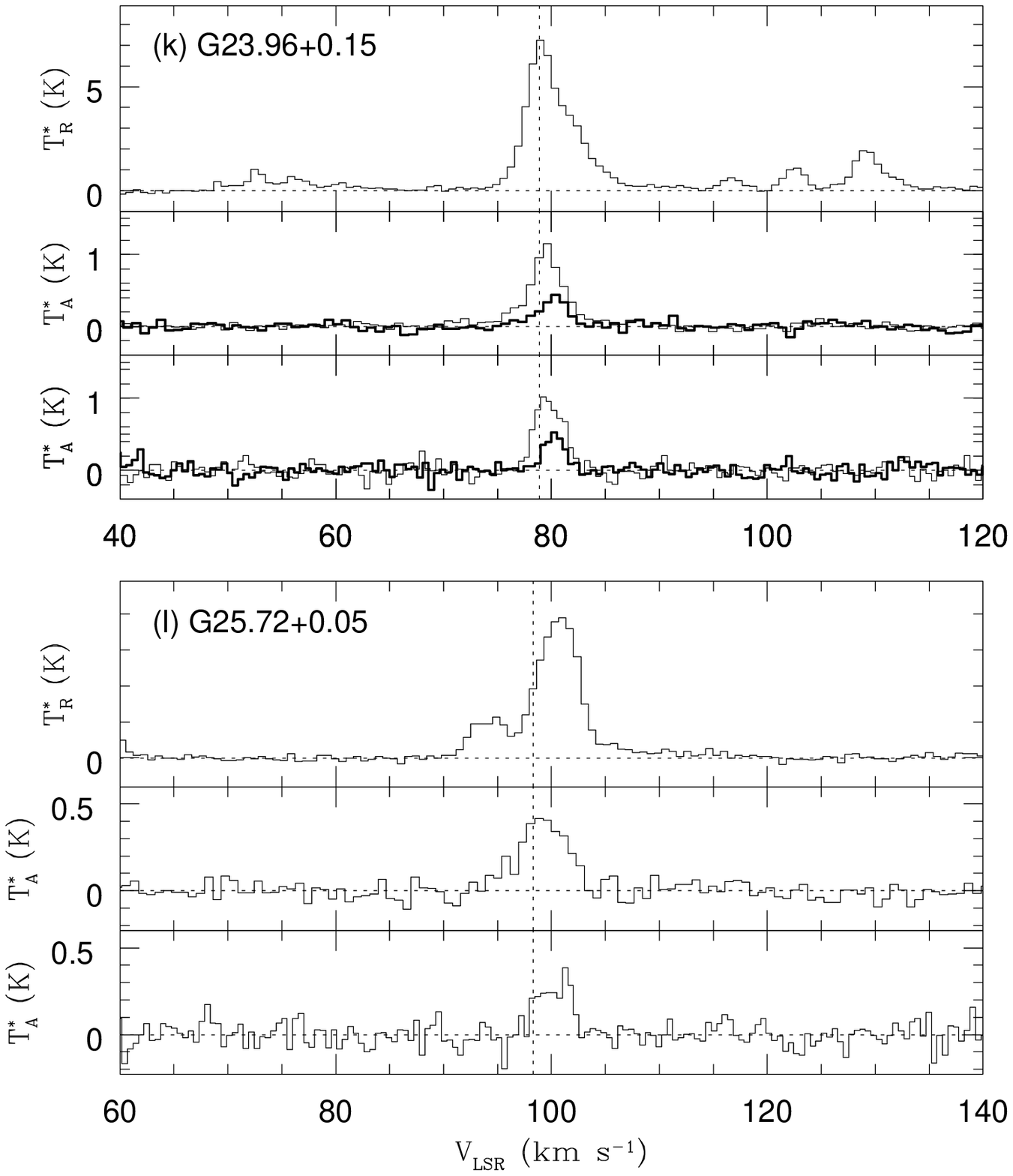}
\end{figure}

\begin{figure}
\vskip 0cm
\figurenum{1}
\epsscale{1.0}
\plottwo{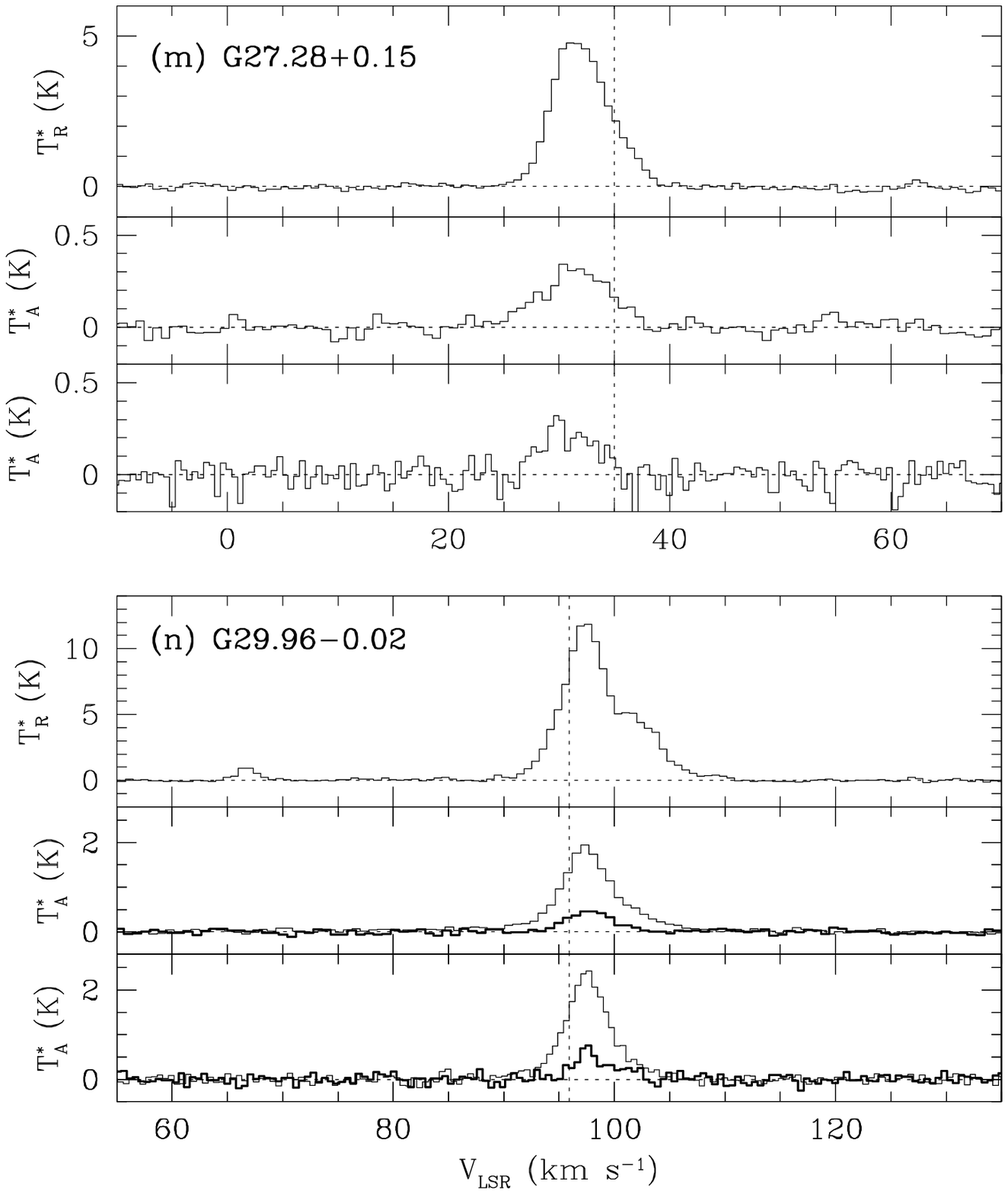}{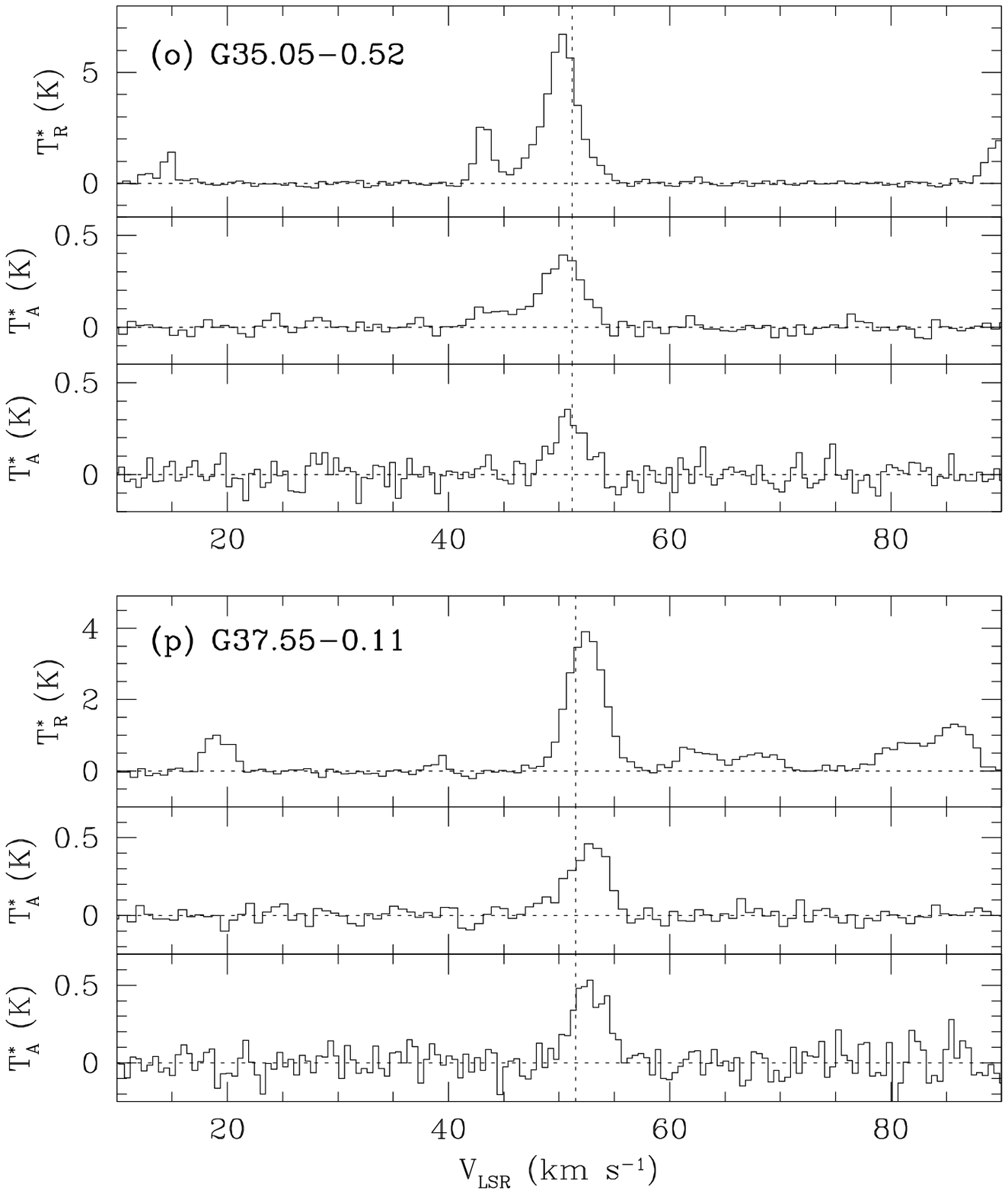}
\center\noindent{Fig. 1.---$Continued$}
\end{figure}


\clearpage

\begin{figure}
\vskip 0cm
\figurenum{2}
\epsscale{1.0}
\end{figure}

\begin{figure}
\vskip 0cm
\figurenum{2}
\epsscale{1.0}
\vskip 0cm
\figcaption{
\cooj\ line integrated intensity ($\int T_{\rm R}^* dv$) maps. 
The integrated velocity range
is presented at the top in each panel, while contour levels and grey
scale flux range are listed at the bottom.
Large crosses represent \uchii\ regions while small crosses indicate 
compact \hii\ regions without \uchii\ regions (\pone).
}
\end{figure}

\begin{figure}
\vskip 0cm
\figurenum{2}
\epsscale{1.0}
\end{figure}
 
\begin{figure}
\vskip 0cm
\figurenum{2}
\epsscale{1.0}
\end{figure}

\begin{figure}
\vskip 0cm
\figurenum{2}
\epsscale{1.0}
\end{figure}

\begin{figure}
\vskip 0cm
\figurenum{2}
\epsscale{1.0}
\end{figure}

\begin{figure}
\vskip 0cm
\figurenum{2}
\epsscale{1.0}
\end{figure}

\begin{figure}
\vskip 0cm
\figurenum{2}
\epsscale{1.0}
\end{figure}


\clearpage

\begin{figure}
\vskip 0cm
\figurenum{3}
\epsscale{1.0}
\plottwo{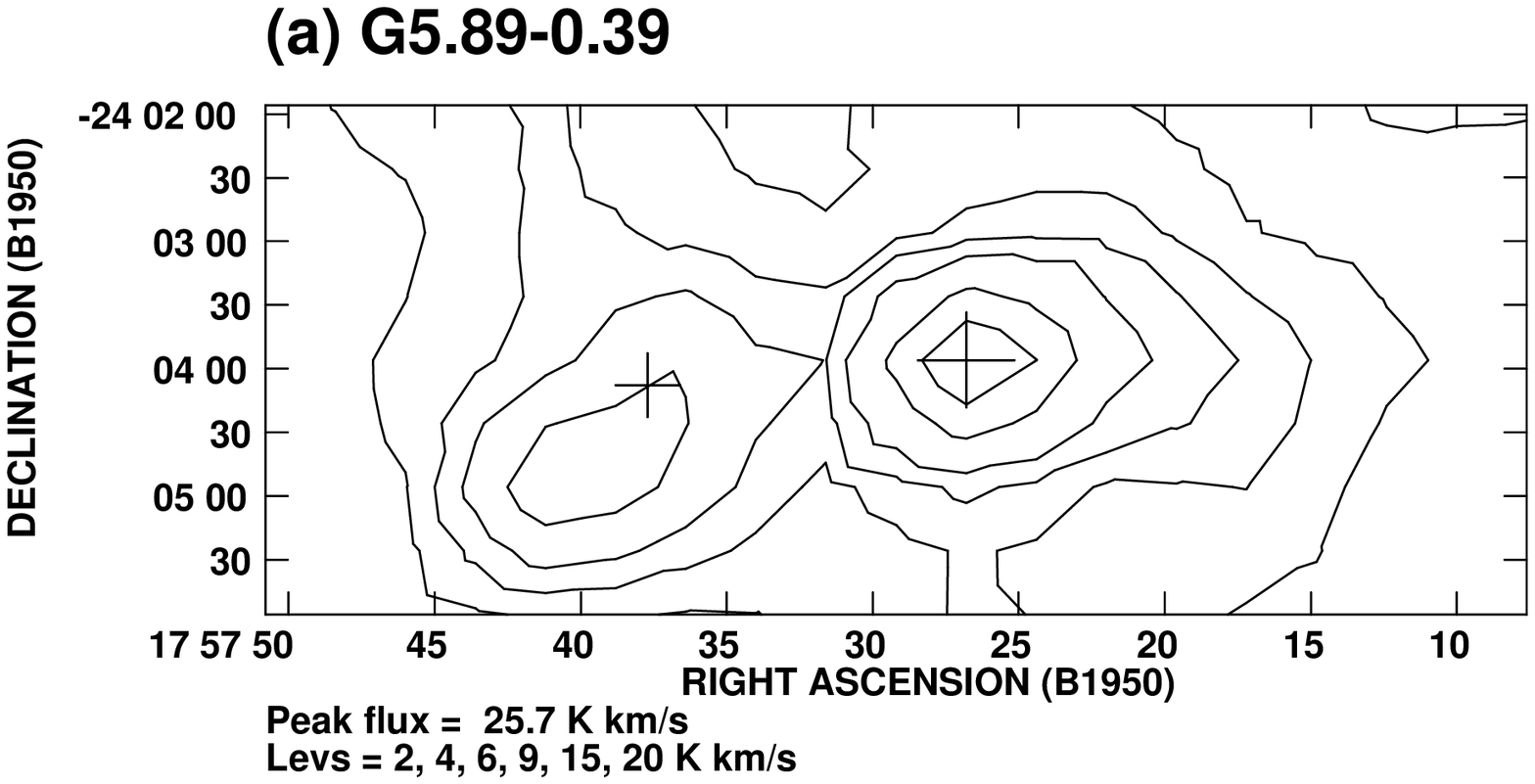}{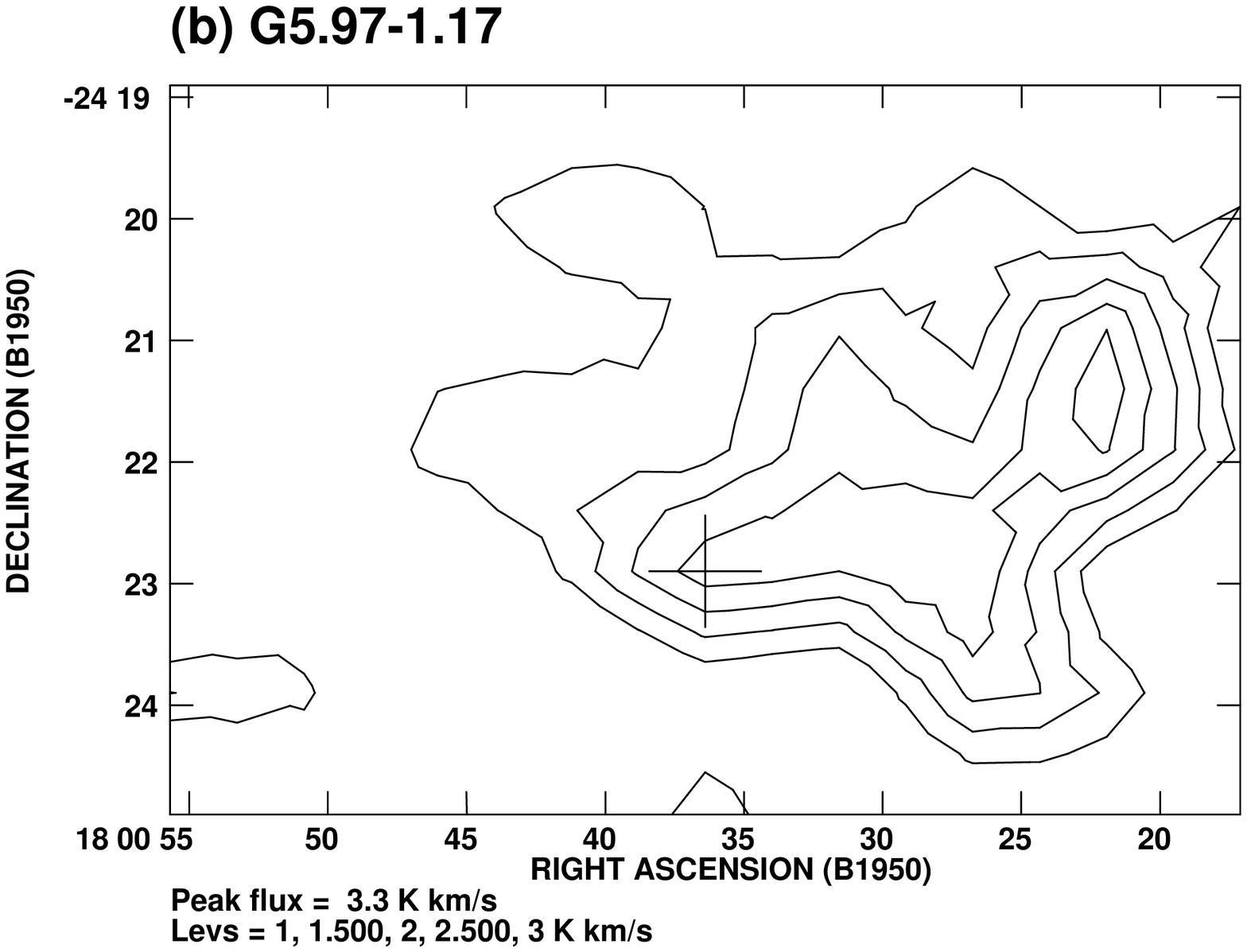}
\end{figure}

\begin{figure}
\vskip 0cm
\figurenum{3}
\epsscale{1.0}
\plottwo{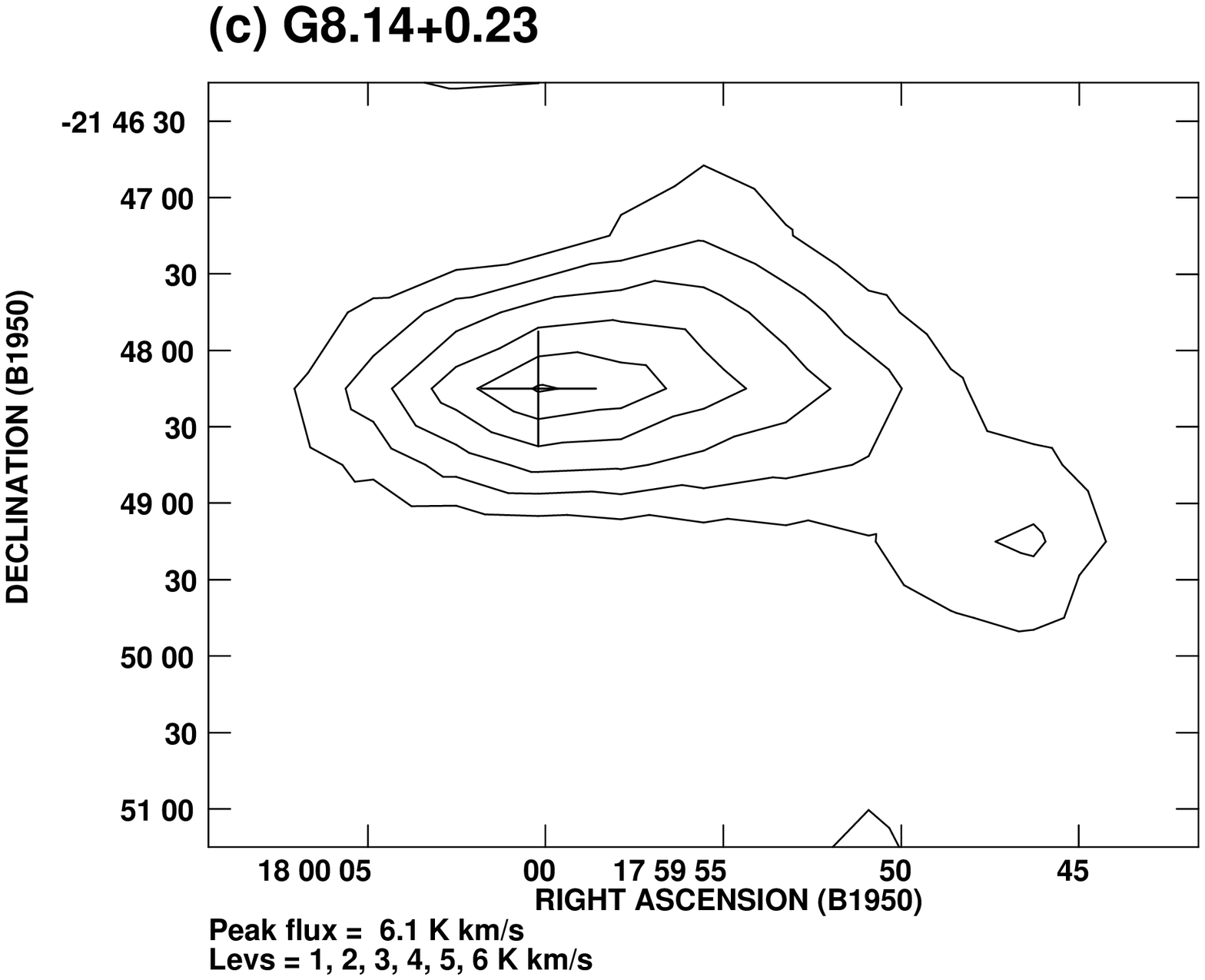}{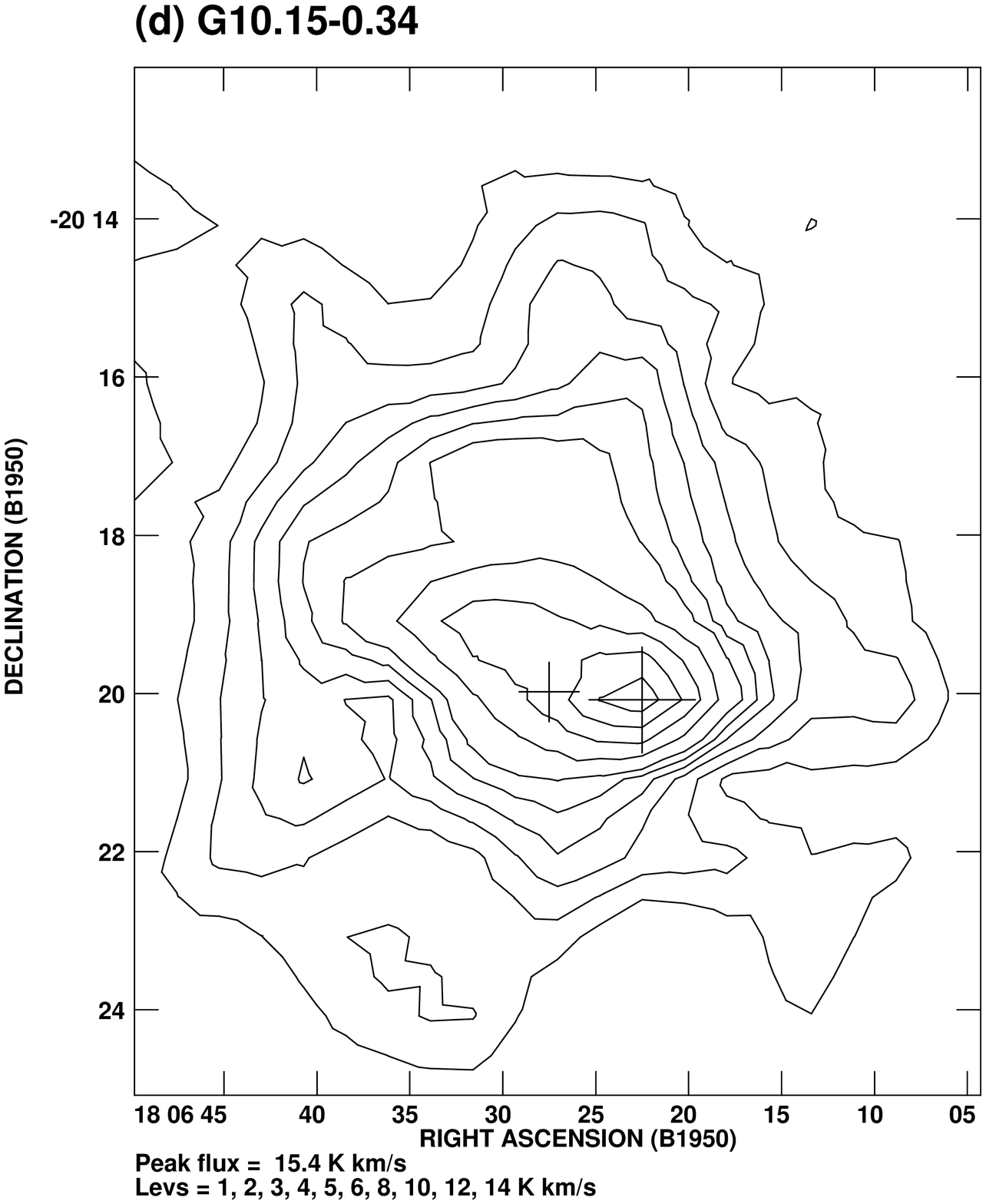}
\vskip 0cm
\figcaption{
\csj\ line integrated intensity ($\int T_{\rm A}^* dv$) maps. 
Contour levels are shown at the
bottom in each panel. Symbols are the same as in Figure 2.
}
\end{figure}

\begin{figure}
\vskip 0cm
\figurenum{3}
\epsscale{1.0}
\plottwo{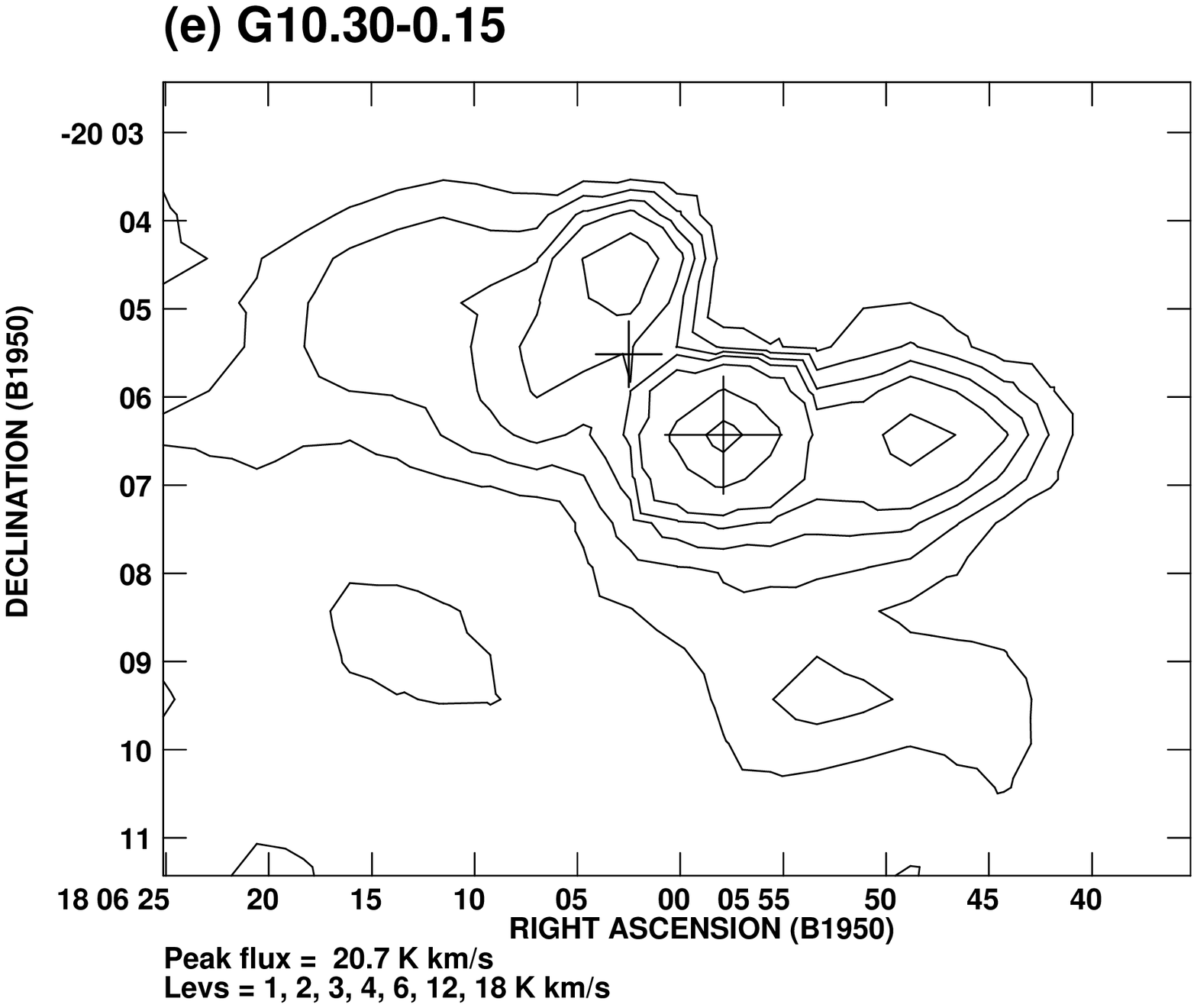}{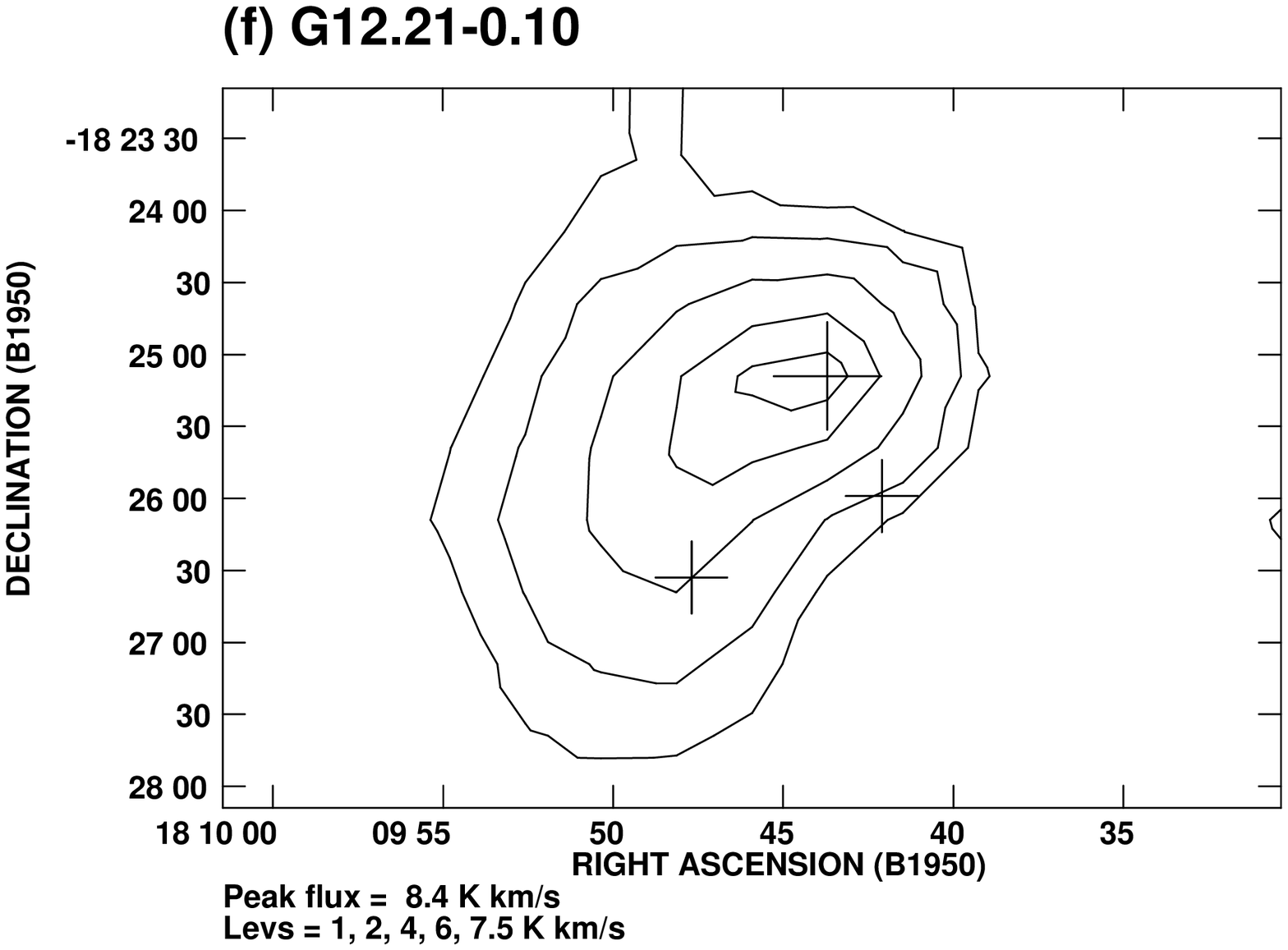}
\end{figure}
 
\begin{figure}
\vskip 0cm
\figurenum{3}
\epsscale{1.0}
\plottwo{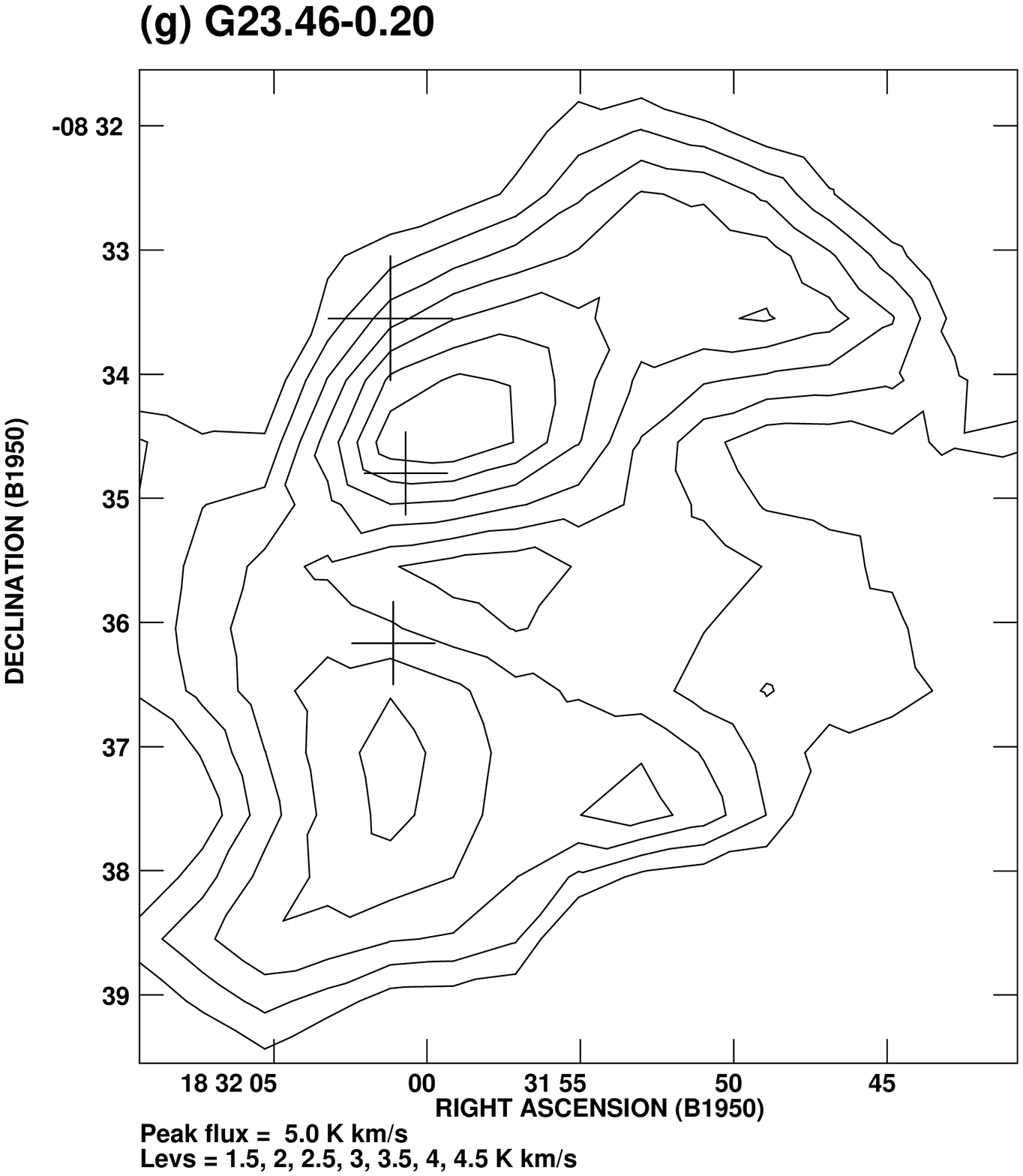}{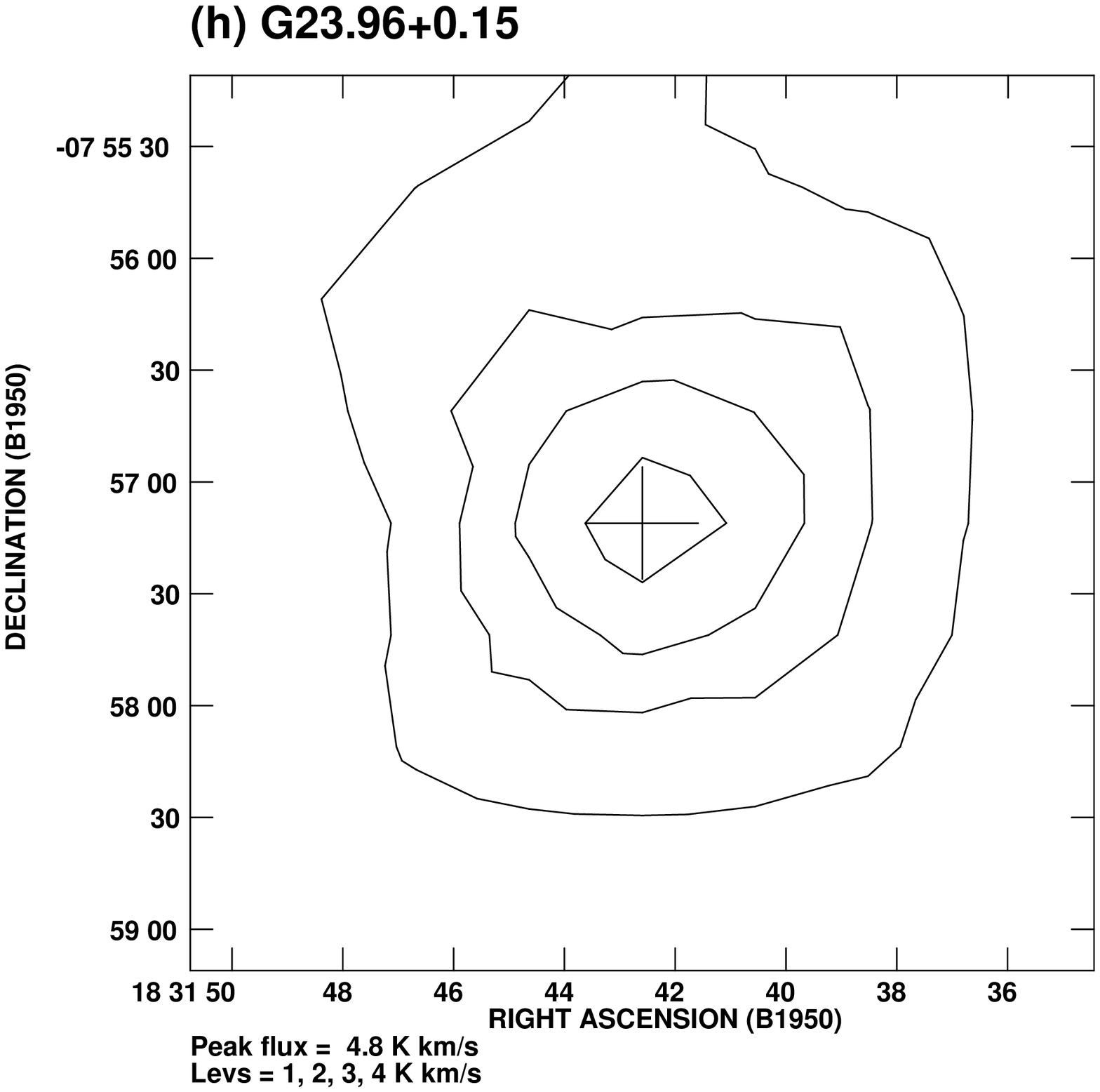}
\center\noindent{Fig. 3.---$Continued$}
\end{figure}

\begin{figure}
\vskip 0cm
\figurenum{3}
\epsscale{1.0}
\plottwo{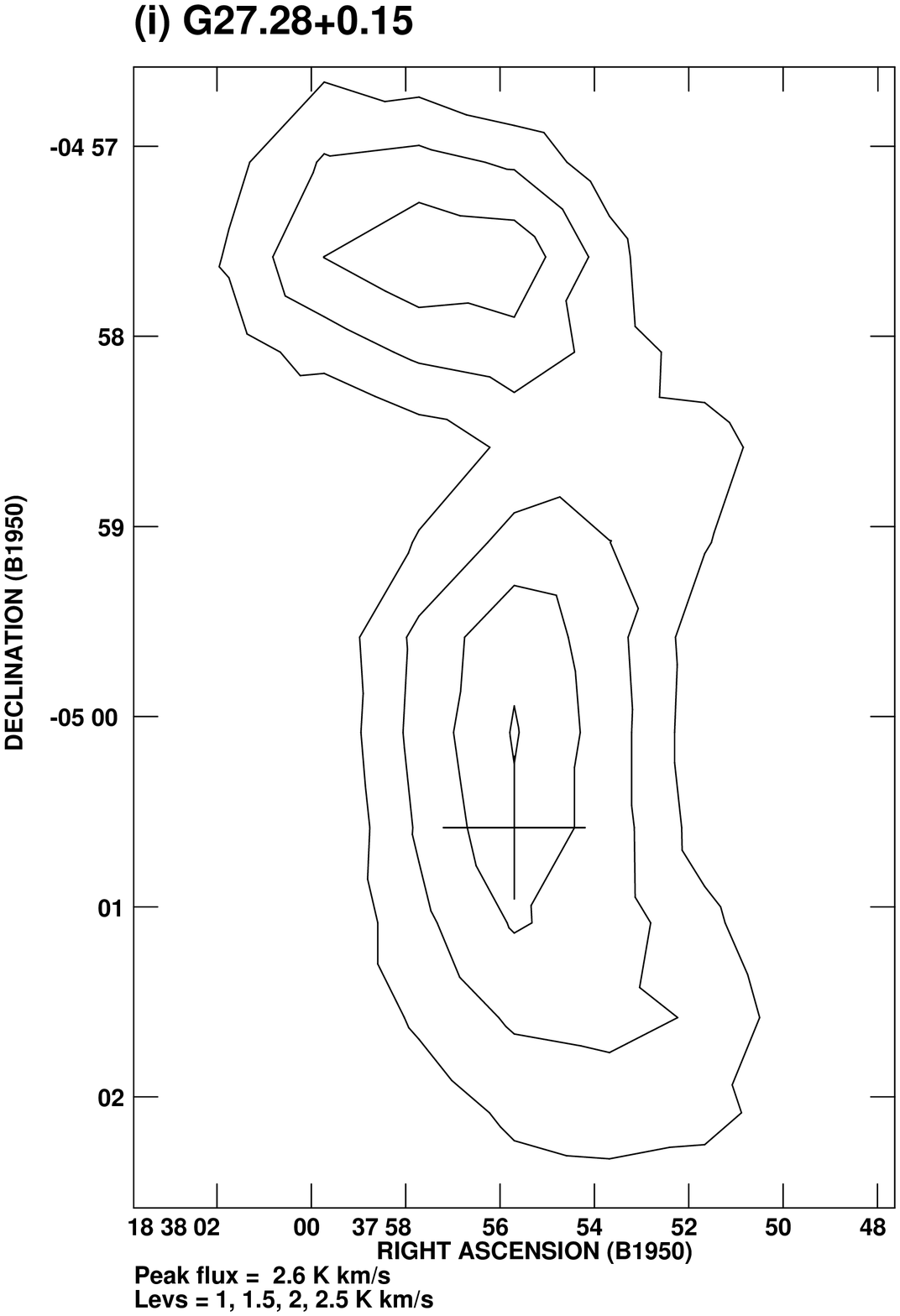}{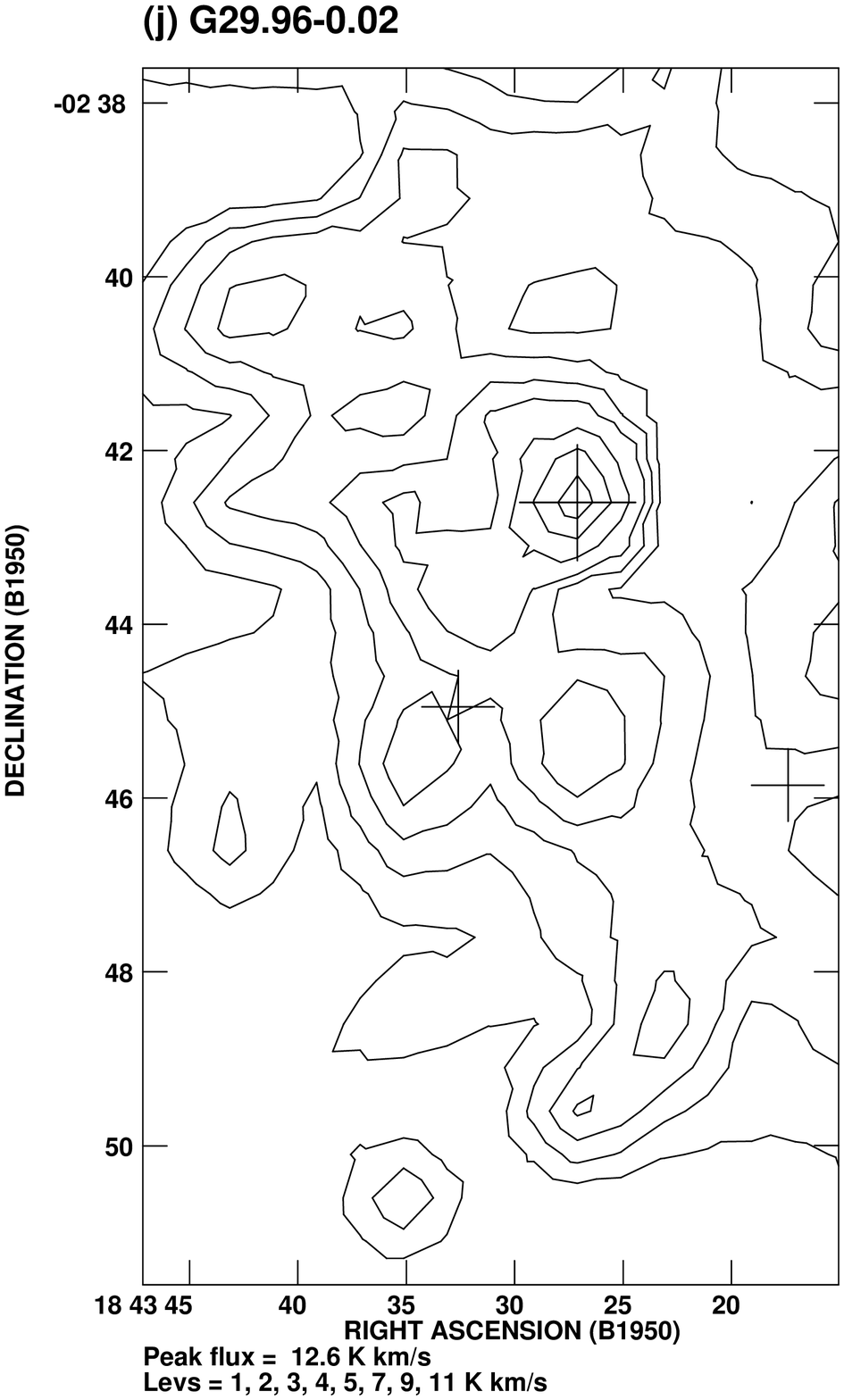}
\center\noindent{Fig. 3.---$Continued$}
\end{figure}

\clearpage
\begin{figure}
\vskip 0cm
\figurenum{4}
\epsscale{1.0}
\plotone{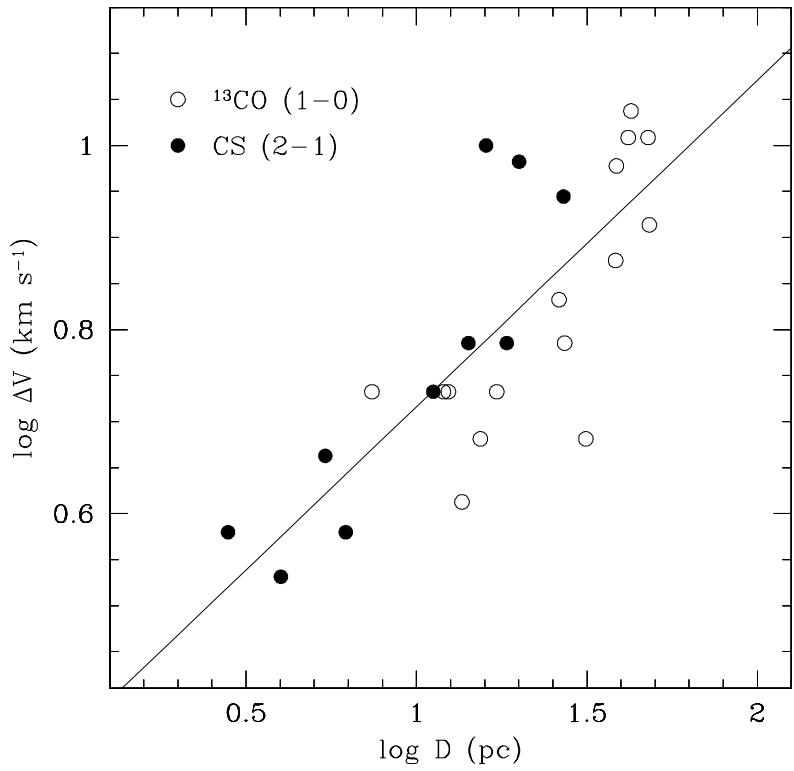}
\vskip 0cm
\figcaption{ 
Plot of \delv\ against $D$.
Open and filled circles indicate the data points
estimated from \coo\ and CS line observations, respectively.
There is a fairly good correlation between the two parameters.
A linear fit to the data points gives
log~\delv\ = 0.35~log~$D$ + 0.4 with a correlation coefficient of 0.8.
The solid line represents the fitted relation.
}
\end{figure}

\clearpage
\begin{figure}
\vskip 0cm
\figurenum{5}
\epsscale{1.0}
\plotone{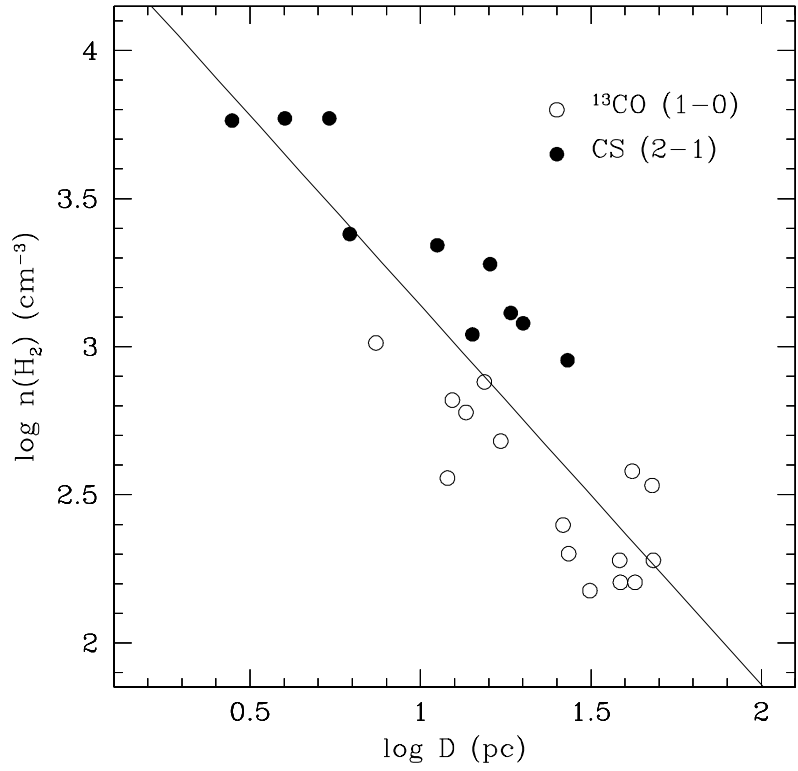}
\vskip 0cm
\figcaption{ 
Comparison of \nhtwo\ with $D$.
A strong correlation is present between the two.
Using a linear-squares fit we obtained
log~\nhtwo\ = $-$1.24 log~$D$ + 4.4, as shown by the solid line.
Symbols are the same as in Figure 4.
}
\end{figure}

\clearpage
\begin{figure}
\vskip 0cm
\figurenum{6}
\epsscale{1.0}
\plotone{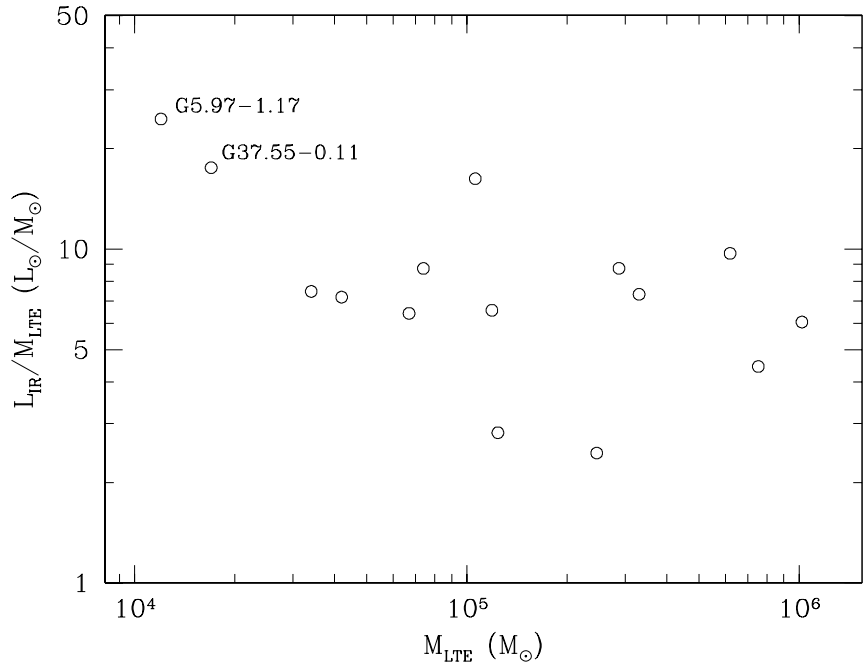}
\vskip 0cm
\figcaption{
The \lir/\mlte\ ratio versus \mlte. We cannot see any apparent correlation
between the two parameters.
}
\end{figure}

\clearpage
\begin{figure}
\vskip 0cm
\figurenum{7}
\epsscale{1.0}
\plotone{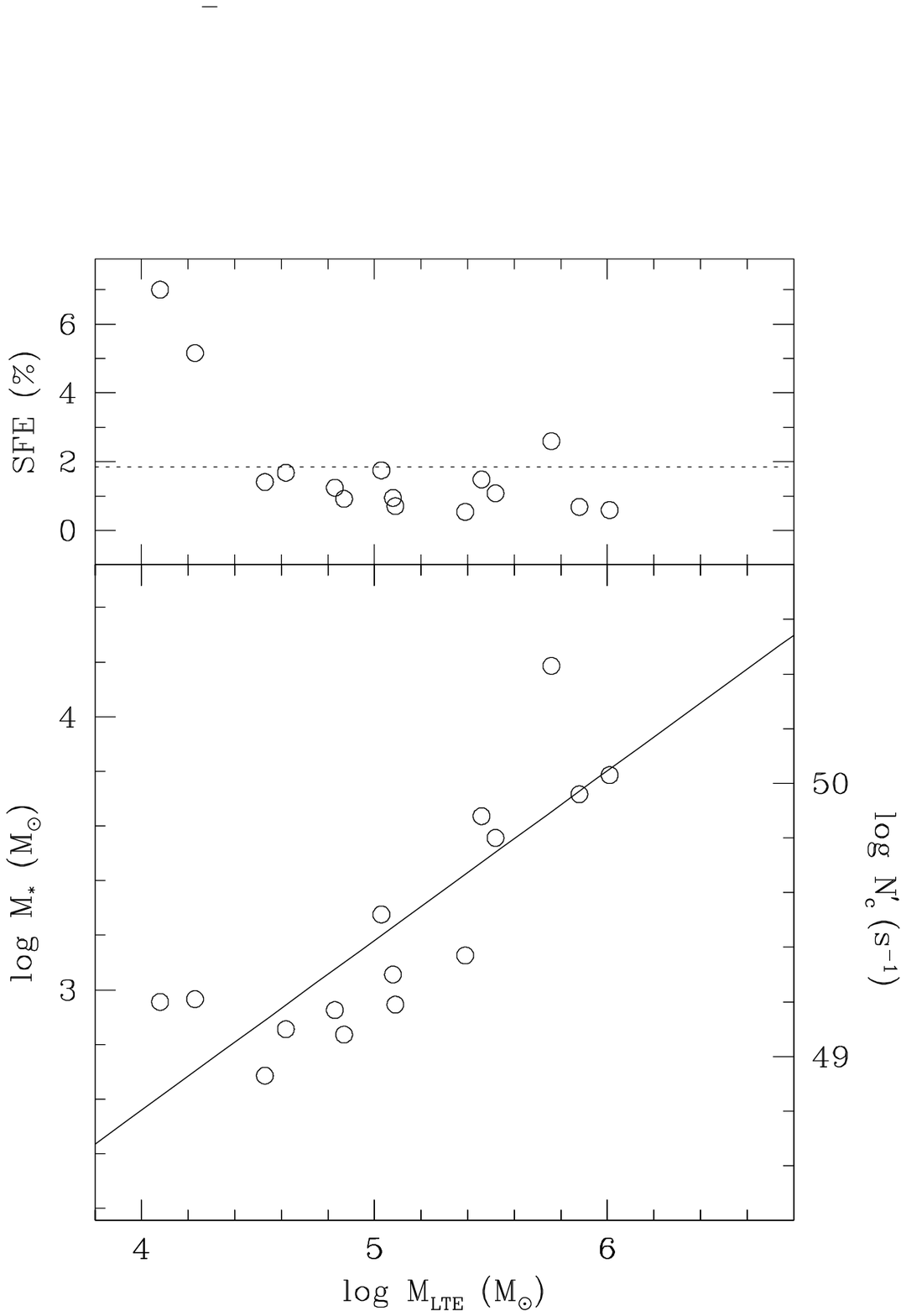}
\vskip 0cm
\figcaption{ 
Stellar mass $M_*$, Lyman continuum photon flux \nc\ (lower panel),
and star formation efficiency (SFE) (upper panel) are compared with
the cloud mass \mlte.
The solid line is the fitted relation in the lower panel.
In the upper panel the dotted line indicates the average value of SFE,
1.9\%. 
}
\end{figure}


\clearpage

\begin{figure}
\vskip 0cm
\figurenum{8}
\epsscale{1.0}
\end{figure}

\begin{figure}
\vskip 0cm
\figurenum{8}
\epsscale{1.0}
\vskip 0cm
\figcaption{
\coo\ integrated intensity image (grey scale) and 21~cm radio continuum image
(contour), made with the VLA DnC-array 
at about 40$''$$\times$20$''$ angular resolution (\pone).
Grey scale flux ranges are the same as in Figure 2.
Contour levels are listed at the bottom in each panel.
Crosses represent the positions of \uchii\ regions.
We can see in most sources that
the UC and compact components of \hii\ regions are associated with
the dense cores of molecular clouds, while
the diffuse extended envelopes often develop in the direction of
decreasing molecular gas density
(see the text for details).
}
\end{figure}

\begin{figure}
\vskip 0cm
\figurenum{8}
\epsscale{1.0}
\end{figure}

\begin{figure}
\vskip 0cm
\figurenum{8}
\epsscale{1.0}
\end{figure}

\begin{figure}
\vskip 0cm
\figurenum{8}
\epsscale{1.0}
\end{figure}
 
\begin{figure}
\vskip 0cm
\figurenum{8}
\epsscale{1.0}
\end{figure}

\begin{figure}
\vskip 0cm
\figurenum{8}
\epsscale{1.0}
\end{figure}
 
\begin{figure}
\vskip 0cm
\figurenum{8}
\epsscale{1.0}
\end{figure}

\clearpage
\begin{figure}
\vskip 0cm
\figurenum{9}
\epsscale{1.0}
\plotone{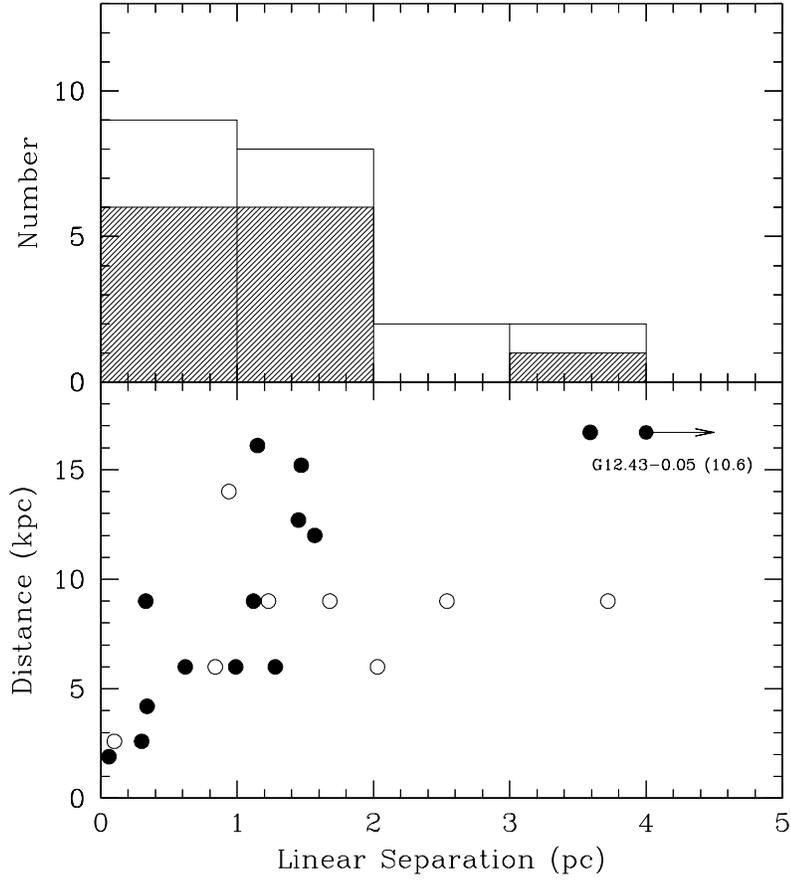}
\vskip 0cm
\figcaption{ 
The projected linear separation between radio continuum peak
and the nearest \coo\ peak.
The lower panel exhibits a plot of the separation versus the distance.
Here filled and open circles are compact \hii\ regions with and without
\uchii\ regions, respectively.
The upper panel is a histogram showing the number of compact \hii\ regions
at each separation interval. Shaded blocks represent the number of
compact \hii\ regions with \uchii\ regions, while open blocks 
indicate the number of those without \uchii\ regions.
}
\end{figure}

\clearpage

\begin{deluxetable}{lcccccccc}
\footnotesize
\tablewidth{0pt}
\tablecaption{\cooj\ Line Observational Parameters}
\tablehead
{
 & & & \multicolumn{2}{c}{Map Center} & & 
\colhead{Map Size} & \colhead{OFFs\tablenotemark{a}} &
\colhead{Adopted\tablenotemark{b}}
\\ \cline{4-5} \cline{7-7}
 & \colhead{$\alpha$(1950)} & \colhead{$\delta$(1950)} & 
 \colhead{$\alpha$(1950)} & \colhead{$\delta$(1950)} & & 
\colhead{$\theta_\alpha \times \theta_\delta$} &
\colhead{($l$, $b$)} & 
\colhead{Distance}
\\
\colhead{Source} & 
\colhead{($^{\rm h}~ ^{\rm m}~ ^{\rm s}$)} & \colhead{($^\circ~ '~ ''$)} & 
\colhead{($^{\rm h}~ ^{\rm m}~ ^{\rm s}$)} & \colhead{($^\circ~ '~ ''$)} & & 
\colhead{($'$)} & \colhead{($^\circ$)} & \colhead{(kpc)} 
}
\startdata
G5.89$-$0.39 ..... & 17 57 26.8 & $-$24 03 56 & 17 57 37.7 & $-$24 03 51 & & 
	$20 \times 20$ & ~(5.00, $-$2.00) & ~2.6 \\
G5.97$-$1.17 ..... & 18 00 36.4 & $-24$ 22 54 & 18 00 36.4 & $-24$ 22 54 & & 
	$20 \times 20$ & ~(5.90, $-$1.50) & ~1.9 \\
G6.55$-$0.10 ..... & 17 57 47.4& $-23$ 20 30 & 17 57 47.4& $-23$ 20 30 & &
	$10 \times 10$ & ~(6.55, +1.50) & 16.7\\
G8.14+0.23 ..... & 18 00 00.2& $-21$ 48 15 & 18 00 00.2& $-21$ 48 15 & &
	$20 \times 15$ & ~(9.00, +0.50) & ~4.2\\
G10.15$-$0.34 ... & 18 06 22.5& $-$20 20 05 & 18 06 09.1& $-$20 09 23 & &
	$30 \times 40$ & (10.15, $-$1.50) & ~6.0\\
G10.30$-$0.15 ... & 18 05 57.9& $-$20 06 26 & $''$ & $''$ &
	$''$ & $''$ & $''$ & ~6.0\\
G12.21$-$0.10 ... & 18 09 43.7& $-18$ 25 09 & 18 09 43.7& $-18$ 25 09 & &
	$20 \times 10$ & (12.00, +1.00) & 16.1\\
G12.43$-$0.05 ... & 18 09 58.9& $-18$ 11 58 & 18 09 58.9& $-18$ 11 58 & &
	$~7 \times ~7$ & (12.00, +1.00) & 16.7\\
G23.46$-$0.20 ... & 18 32 01.2& $-08$ 33 33 & 18 32 00.0& $-08$ 35 00 & &
	$15 \times 15$ & (22.86, $-$0.86) & ~9.0\\
G23.71+0.17 ... & 18 31 10.3& $-08$ 09 36 & 18 31 10.3& $-08$ 09 36 & &
	$~9 \times ~7$ & (23.87, +0.83) & ~9.0\\
G23.96+0.15 ... & 18 31 42.6& $-07$ 57 11 & 18 31 42.6& $-07$ 57 11 & &
	$~9 \times ~7$ & (23.87, +0.83) & ~6.0\\
G25.72+0.05 ... & 18 35 21.6& $-06$ 26 27 & 18 35 28.0& $-06$ 27 00 & &
	$10 \times ~7$ & (25.66, $-$1.00) & 14.0\\
G27.28+0.15 ... & 18 37 55.7& $-05$ 00 35 & 18 37 55.7& $-05$ 00 35 & &
	$10 \times ~7$ & (26.94, $-$0.50) & 15.2\\
G29.96$-$0.02 ... & 18 43 27.1& $-02$ 42 36 & 18 43 30.8& $-02$ 45 10 & &
	$20 \times 15$ & (30.00, +0.50) & ~9.0\\
G35.05$-$0.52 ... & 18 54 37.1& +01 35 01 & 18 54 37.1& +01 35 01 & &
	$13 \times ~7$ & (34.10, $-$1.00) & 12.7\\
G37.55$-$0.11 ... & 18 57 46.8& +03 59 00 & 18 57 46.8& +03 59 00 & &
	$~7 \times ~7$ & (37.55, $-$0.50) & 12.0 
\enddata
\tablenotetext{a}{Reference positions with some \coo\ emission (see the text).}
\tablenotetext{b}{From references given by Wood \& Churchwell (1989).}
\end{deluxetable}

\clearpage

\begin{deluxetable}{lccccc}
\footnotesize
\tablewidth{0pt}
\tablecaption{\cooj\ Line Observational Result}
\tablehead{
  & \colhead{$W_\circ$} & \colhead{$R$} & 
\colhead{$M_{\rm LTE}$} & \colhead{\nhtwo} & \colhead{\delv\tablenotemark{a}} 
\\
\colhead{Source} & \colhead{(K~\kms)} & \colhead{(pc)} & 
\colhead{(10$^4$~\msol)} & \colhead{(10$^2$~\cm)} & \colhead{(\kms)} 
}
\startdata
G5.89$-$0.39 ..... & 15 & ~7.7 & ~~7.4 & ~7.6 & ~4.8 \\ 
G5.97$-$1.17 ..... & 10 & ~3.7 & ~~1.2 & 10.3 & ~4.8 \\ 
G6.55$-$0.10 ..... & 10 & 21.4 & ~33.0 & ~1.6 & ~9.9 \\ 
G8.14+0.23 .....   & 10 & ~6.8 & ~~4.2 & ~6.0 & ~4.1 \\ 
G10.15$-$0.30 ...  & 10 & 24.1 & ~57.5 & ~1.9 & ~8.2 \\ 
G12.21$-$0.10 ...  & 10 & 19.2 & ~28.7 & ~1.9 & ~7.5 \\ 
G12.43$-$0.05 ...  & 10 & 15.7 & ~12.4 & ~1.5 & ~4.8 \\ 
G23.46$-$0.20 ...  & 15 & 20.9 & ~75.4 & ~3.8 & 10.2 \\ 
G23.71+0.17 ...    & 15 & ~8.6 & ~~6.7 & ~4.8 & ~5.8 \\ 
G23.96+0.15 ...    & 10 & ~6.2 & ~~3.4 & ~6.6 & ~5.1 \\ 
G25.72+0.05 ...    & 10 & l3.6 & ~10.6 & ~2.0 & ~5.4 \\ 
G27.28+0.15 ...    & 10 & 13.1 & ~11.9 & ~2.5 & ~6.8 \\ 
G29.96$-$0.02 ...  & 20 & 23.9 & 102.0 & ~3.4 & 10.2 \\ 
G35.05$-$0.52 ...  & 10 & 19.3 & ~24.6 & ~1.6 & ~9.5 \\ 
G37.55$-$0.11 ...  & 10 & ~6.0 & ~~1.7 & ~3.6 & ~4.8 
\enddata
\tablenotetext{a}{FWHM of the average spectrum of the entire mapping area.}
\end{deluxetable}

\clearpage

\begin{deluxetable}{lcccccccccc}
\scriptsize
\tablewidth{0pt}
\tablecaption{CS Line Parameters}
\tablehead{
  & \colhead{OFFs}  &
\multicolumn{5}{c}{CS (2$-$1)} & & \multicolumn{3}{c}{CS (3$-$2)}
\\ \cline{3-7} \cline{9-11}
  & \colhead{($l$, $b$)}  & 
\colhead{\vlsr} & \colhead{\ta} & \colhead{\delv} & 
\colhead{$\int T_{\rm A}^* dv$} & \colhead{\Ncs}  &&
\colhead{\vlsr} & \colhead{\ta} & \colhead{\delv}
\\
\colhead{Source} & \colhead{($^\circ$)} &
\colhead{(\kms)} & \colhead{(K)} & \colhead{(\kms)} &
\colhead{(K~\kms)} & \colhead{($10^{13}$~cm$^{-2}$)} &&
\colhead{(\kms)} & \colhead{(K)} & \colhead{(\kms)}
}
\startdata
G5.89$-$0.39 ..... & ~(5.00, $-$1.00) &
	~~9.6 & 3.80 & 6.1 & 25.9 & 44.0 && ~~9.8 & 4.30 & 5.6 \\
G5.97$-$1.17 ..... & ~(5.00, $-$1.00) & 
	~~9.6 & 0.73 & 3.4 & ~2.8 & ~4.7 && ~~9.8 & 0.95 & 2.3 \\
G6.55$-$0.10 ..... & ~(7.00, $-$1.00) & 
	~15.0 & 0.33 & 6.9 & ~2.6 & ~4.4 && ~14.3 & 0.30 & 6.6 \\
G8.14+0.23 ..... & ~(8.60, +0.60) & 
	~19.6 & 0.85 & 6.1 & ~6.2 & 10.4 && ~19.8 & 0.67 & 5.1 \\
G10.15$-$0.34 ... & (10.15, +0.50) & 
	~~9.6 & 1.31 & 8.8 & 16.6 & 28.1 && ~10.8 & 1.05 & 7.1 \\
G10.30$-$0.15 ... & (10.15, +0.50) & 
	~13.5 & 1.90 & 8.8 & 21.8 & 37.0 && ~13.8 & 2.00 & 7.7 \\
G12.21$-$0.10 ... & (12.21, +0.50) & 
	~25.0 & 0.98 & 8.0 & ~8.4 & 14.2 && ~24.9 & 0.91 & 6.4 \\
G12.43$-$0.05 ... & (12.21, +0.50) & 
	~19.6 & 0.28 & 3.1 & ~1.2 & ~2.0 && \nodata & $<$0.1~ & \nodata \\
G23.46$-$0.20 ... & (23.46, +0.50) & 
	101.9 & 0.30 & 7.7 & ~2.8 & ~4.7 && 102.8 & 0.17 & 7.1 \\
G23.71+0.17 ... & (23.80, +0.50) & 
	114.6 & 0.70 & 6.1 & ~5.1 & ~8.7 && 114.2 & 0.60 & 5.6 \\
G23.96+0.15 ... & (23.80, +0.50) & 
	~79.6 & 1.15 & 2.7 & ~4.8 & ~8.2 && ~79.2 & 1.02 & 3.6 \\
G25.72+0.05 ... & (25.65, $-$0.60) & 
	~98.9 & 0.42 & 5.4 & ~2.3 & ~3.9 && 101.3 & 0.28 & 4.1 \\
G27.28+0.15 ... & (27.30, $-$0.50) & 
	~30.4 & 0.34 & 5.4 & ~2.4 & ~4.2 && ~29.8 & 0.30 & 4.6 \\
G29.96$-$0.02 ... & (30.00, +0.50) & 
	~97.3 & 1.94 & 5.0 & 12.4 & 21.1 && ~97.7 & 2.41 & 4.1 \\
G35.05$-$0.52 ... & (35.15, $-$1.00) & 
	~51.1 & 0.40 & 3.8 & ~2.l & ~3.6 && ~50.8 & 0.35 & 2.8 \\
G37.55$-$0.11 ... & (37.55, $-$0.50) &
	~52.7 & 0.46 & 4.2 & ~2.0 & ~3.5 && ~52.8 & 0.53 & 3.6 
\enddata
\end{deluxetable}

\clearpage
 
\begin{deluxetable}{lccccccccccccc}
\footnotesize
\tablewidth{0pt}
\tablecaption{\css\ Line Parameters}
\tablehead{
  & \multicolumn{4}{c}{\css\ (2$-$1)} && \multicolumn{4}{c}{\css\ (3$-$2)}
\\ \cline{2-5} \cline{7-10}
  & \colhead{\vlsr} & \colhead{\ta} & \colhead{\delv} & $\tau_{\rm p}$ &&
    \colhead{\vlsr} & \colhead{\ta} & \colhead{\delv} & $\tau_{\rm p}$ &
\\
\colhead{Source} &
\colhead{(\kms)} & \colhead{(K)} & \colhead{(\kms)} & &&
\colhead{(\kms)} & \colhead{(K)} & \colhead{(\kms)} &
}
\startdata
G5.89$-$0.39 ..... &
	~~9.6 & 0.70 & 5.4& 0.26 && ~~9.2 & 1.03 & 3.1 & 0.35\\ 
G10.15$-$0.34 ... &
	~10.4 & 0.32 & 3.1 & 0.36 && ~10.3 & 0.50 & 2.6 & 0.88\\
G10.30$-$0.15 ... &
	~15.1 & 0.40 & 7.0 & 0.30 && 13.4 & 0.62 & 2.1 & 0.48 \\ 
G12.21$-$0.10 ... &
     ~23.5 & 0.25 & 4.7 & 0.38 && \nodata & $<$0.10~~ & \nodata & $<$0.14~~~\\ 
G23.96+0.15 ... &
	~80.4 & 0.44 & 2.3 & 0.64 && ~80.3 & 0.52 & 2.6 & 0.97\\
G29.96$-$0.02 ... &
	~98.1 & 0.46 & 4.7 & 0.35 && ~97.7 & 0.75 & 2.1 & 0.48\\ 
\enddata
\end{deluxetable}

\clearpage

\begin{deluxetable}{lcccccc}
\footnotesize
\tablewidth{0pt}
\tablecaption{Properties of CS Clouds}
\tablehead{
  & \colhead{$R$} & \colhead{$M_{\rm LTE}$} & 
\colhead{\nhtwo} & \colhead{\delv\tablenotemark{a}} & 
\multicolumn{2}{c}{Fraction\tablenotemark{b}} 
\\ \cline{6-7}
\colhead{Source} & \colhead{(pc)} & \colhead{($10^4$~\msol)} & 
\colhead{($10^3$~cm$^{-3}$)} & \colhead{(\kms)} & 
\colhead{Area (\%)} & \colhead{Mass (\%)}
}
\startdata
G5.89$-$0.39 ..... & ~2.7 & ~2.6 & 5.9 & ~4.6 & 12 & 35\\
G5.97$-$1.17 ..... & ~1.4 & ~0.3 & 5.8 & ~3.8 & 14 & 25\\
G8.14+0.23 ..... & ~2.0 & ~1.0 & 5.9 & ~3.4 & ~9 & 24\\
G10.15$-$0.34 ... & ~8.0 &  21.3 & 1.9 & 10.0 & 16 & 48\\
G10.30$-$0.15\tablenotemark{c} ... & ~5.6 &  ~8.2 & 2.2 & ~5.4 & $''$ & $''$\\
G12.21$-$0.10 ... & ~9.2 &  21.4 & 1.3 & ~6.1 & 23 & 75\\
G23.46$-$0.20 ... & 10.0 & 25.1 & 1.2 & ~9.6 & 23 & 33\\
G23.96+0.15 ... & ~3.1 & ~1.6 & 2.4 & ~3.8 & 25 & 47\\
G27.28+0.15 ... & ~7.1 & ~8.5 & 1.1 & ~6.1 & 29 & 71\\
G29.96$-$0.02 ... & 13.5 & 47.1 & 0.9 & ~8.8 & 32 & 46\\
\enddata
\tablenotetext{a}{FWHM of the average spectrum of the entire mapping area.}
\tablenotetext{b}{Fraction of CS parameter relative to \coo\ parameter.}
\tablenotetext{c}{Included with G10.15$-$0.34 to calculate the fraction values.}
\end{deluxetable}

\clearpage

\begin{deluxetable}{lcccccc}
\footnotesize
\tablewidth{0pt}
\tablecaption{Star Formation Activity of Molecular Clouds}
\tablehead{
  &  & \colhead{\lir} & 
\colhead{\lir/\mlte} & \colhead{log~\nc\tablenotemark{a}} &
\colhead{$M_\ast$\tablenotemark{b}} & \colhead{SFE}
\\
\colhead{Source} & \colhead{$R(\overline{T_d}, \beta)$} &
\colhead{(10$^5$~\lsol)} & \colhead{(\lsol/\msol)} & 
\colhead{(s$^{-1}$)} & \colhead{(10$^2$~\msol)} & \colhead{(\%)}
}
\startdata
G5.89$-$0.39 ..... & 1.37 &  ~6.5 & ~8.8 & 49.08 & ~~~6.9 & 0.9\\
G5.97$-$1.17 ..... & 1.41 & ~2.9 & 24.5 & 49.20 & ~~~9.0 & 7.0\\
G6.55$-$0.10 ..... & 1.38 & 24.2 & ~7.3 & 49.80 & ~36.0 & 1.1\\
G8.14+0.23 .....   & 1.37 & ~3.0 & ~7.2 & 49.10 & ~~~7.2 & 1.7\\
G10.15$-$0.30 ...  & 1.36 & 60.2 & ~9.7 & 50.43 & 153.4 & 2.6 \\
G12.21$-$0.10 ...  & 1.40 & 25.1 & ~8.8 & 49.88 & ~43.2 & 1.5\\
G12.43$-$0.05 ...  & 1.67 & ~3.5 & ~2.8 & 49.19 & ~~~8.8 & 0.7\\
G23.46$-$0.20 ...  & 1.48 & 33.6 & ~4.5 & 49.96 & ~52.0 & 0.7\\
G23.71+0.17 ...    & 1.36 & ~4.3 & ~6.4 & 49.17 & ~~~8.4 & 1.2\\
G23.96+0.15 ...    & 1.36 & ~2.5 & ~7.5 & 48.97 & ~~~4.9 & 1.4\\
G25.72+0.05 ...    & 1.36 & 17.2 & 16.3 & 49.52 & ~18.9 & 1.7\\
G27.28+0.15 ...    & 1.42 & ~7.8 & ~6.6 & 49.30 & ~11.4 & 0.9\\
G29.96$-$0.02 ...  & 1.36 & 61.7 & ~6.1 & 50.03 & ~61.1 & 0.6 \\
G35.05$-$0.52 ...  & 1.38 & ~6.0 & ~2.5 & 49.37 & ~13.4 & 0.5\\
G37.55$-$0.11 ...  & 1.36 & ~3.0 & 17.5 & 49.21 & ~~~9.2 & 5.2
\enddata
\tablenotetext{a}{Taken from \pone.}
\tablenotetext{b}{See \S~4.1.}
\end{deluxetable}

\end{document}